\documentclass[twocolumn,eqsecnum,tightenlines,floats,floatfix,aps,nofootinbib,preprintnumbers]{revtex4}
\def\bea{\begin{eqnarray}}
\def\eea{\end{eqnarray}}

\def\st#1{{\kern-4pt} \not\!#1}

\def\sp{\kern +3pt}
\def\sm{\kern -3pt}
\def\spQ{\kern +6pt}

\usepackage[usenames]{color}

\def\be{\begin{equation}}
\def\ee{\end{equation}}
\def\ba{\begin{eqnarray}}
\def\ea{\end{eqnarray}}

\usepackage{graphics}
\usepackage{graphicx}
\usepackage{epsf} 
\usepackage{amsmath}
\usepackage{amssymb}
\usepackage{slashed}
\usepackage{bm}

\begin{document}

\phantom{0}
\vspace{-0.2in}
\hspace{5.5in}

\preprint{ {\bf LFTC-18-7/28} }

\vspace{-1in}%\parbox{1.5in}{ \vspace{-9.6in}}  % moves the preprint box down

\title
{\bf Holographic estimate of the meson cloud 
contribution to nucleon axial form factor}
\author{G.~Ramalho$^{1,2}$ 
\vspace{-0.1in} }

\affiliation{$^1$Laborat\'orio de F\'{i}sica Te\'orica e Computacional -- LFTC,
Universidade Cruzeiro do Sul, 01506-000, S\~ao Paulo, SP, Brazil
\vspace{-0.15in}}

\affiliation{$^2$International Institute of Physics,
Federal University of Rio Grande do Norte,
Campus Universit\'ario - Lagoa Nova  CP.~1613,
Natal, Rio Grande do Norte 59078-970, Brazil}

\vspace{0.2in}
\date{\today}

\phantom{0}

\begin{abstract}
We use light-front Holography to estimate 
the valence quark and the meson cloud contributions 
to the nucleon axial form factor.
The free couplings of the holographic model 
are determined by the empirical data and 
by the information extracted from lattice QCD.
The holographic model provides a good description 
of the empirical data when we consider a 
meson cloud mixture of about 30\% in the physical nucleon state.
The estimate of the valence quark contribution 
to the nucleon axial form factor  
compares well with the lattice QCD data for small pion masses.
Our estimate of the meson cloud contribution 
to the nucleon axial form factor has 
a slower falloff  with the square  momentum transfer 
compared to typical estimates from 
quark models with meson cloud dressing.
\end{abstract}

%\phantom{0}
%\vspace{7.0in}
%\vspace{-6in}
\vspace*{0.9in}  % sets how far the title is below the preprint box
\maketitle

\section{Introduction}
\label{secIntro}

In recent years, it was 
found that the combination of the 5D gravitational
anti-de Sitter (AdS) space and conformal field theories (CFT) 
can be used to study QCD in the 
confining regime~\cite{Maldacena99,Witten98,Gubser98,Brodsky15}.
Using this formalism one can  relate the results 
from AdS/CFT with the results from light-front dynamics 
based on a Hamiltonian that include 
the confining mechanism of QCD (AdS/QCD)~\cite{Brodsky15}.
In the limit of massless quarks,
one can relate the AdS holographic variable $z$ 
with the impact separation $\zeta$, which measures the distance of 
constituent partons inside the hadrons~\cite{Brodsky15,Brodsky08a,Teramond09a}.
This correspondence (duality) between the two formalisms is known 
as light-front holography or holographic QCD.

Over the last few years light-front holography
has been used to study several proprieties of the hadrons.
The soft-wall formulation of the  
light-front holography introduces a holographic mass scale $\kappa$,
which is fundamental for the description of the hadron spectrum 
(mesons and baryons) and  hadron wave 
functions~\cite{Brodsky15,Karch06,Grigoryan07a,Branz10a,Gutsche12b,Teramond15a}.
This scale can be estimated from the holographic expression for 
the $\rho$ mass 
$m_\rho \simeq 2 \kappa$~\cite{Brodsky15,Grigoryan07a}.
Examples of applications of light-front holography
are in the calculation of parton distribution functions,
hadron structure form factors among others~\cite{Brodsky15,Brodsky08a,Chakrabarti13a,Abidin09,Teramond11a,Liu15c,Sufian17a,Gutsche13a,Gutsche12a,RoperHol,RoperAn,Vega11,Chakrabarti13}.

In the light-front formalism one can represent the 
wave functions of the hadrons using an expansion of Fock states 
with a well defined number of partons~\cite{Brodsky15}.
In the case of baryons, the first 
term corresponds to the three-quark state ($qqq$).
The following terms are excitations associated 
with a gluon, $(qqq)g$, with a quark-antiquark pair, 
$(qqq)q \bar q$, and higher order terms.
Those states can be labeled in terms 
of the number of partons $\tau=3,4,5,...$, respectively. 
The calculation of structure form 
factors between baryon states can then be performed 
using the light-front wave functions 
and the interaction vertices associated with 
the respective transition~\cite{Abidin09,Gutsche12a,RoperHol}.
The form factors can also be  
expanded in contributions 
from the valence quarks and 
in contributions from the meson cloud~\cite{Brodsky15,Gutsche12a,Gutsche13a}.
Examples of calculations of the nucleon and the  
nucleon to Roper electromagnetic form factors
can be found in Refs.~\cite{Gutsche12a,RoperHol,RoperAn,Brodsky15,Teramond11a,Abidin09,Chakrabarti13a,Liu15c,Sufian17a,Gutsche13a}.

In principle the leading twist approximation,
associated with the three-quark state, is sufficient 
to explain the dominant contribution of the 
form factors related to the electromagnetic transitions 
between baryon states, 
particularly at large momentum transfer.
In the case of the nucleon and the Roper,
the electromagnetic form factors can be 
described in a good approximation by the 
valence quark effects 
(leading twist approximation)~\cite{Brodsky15,RoperHol,Chakrabarti13a,Liu15c}.
There is, however a rising interest in checking 
if the holography can be used to estimate 
higher order corrections to the transition form factors,
particularly, in the corrections associated with 
the meson cloud excitations, 
related in the light-front formalism to 
the state $(qqq)q \bar q$, of order $\tau =5$~\cite{Gutsche12a,Gutsche13a,Sufian17a}.

The question of whether the light-front meson cloud contribution 
is important or not is pertinent, because in principle 
the corrections associated with the meson cloud 
should be expressed in terms of parameters 
related to the microscopic structure, 
such as meson-baryon couplings and the photon-meson 
couplings~\cite{NucleonMC1,NucleonMC2,NucleonMC3}.
As discussed later, 
in the case of a holographic model, the estimates
of the transition form factors depend only on the couplings 
associated with quarks 
without explicit reference to the 
substructure associated with the meson cloud.

In this work we study the axial structure of the nucleon
using a holographic model based on a soft-wall confining potential.
The weak structure of the nucleon is characterized by 
the axial form factor, $G_A$, and the induced pseudoscalar form factor, $G_P$.
The study of the nucleon axial structure is important 
because it provides complementary information on the well known electromagnetic structure 
and also because involves both strong and weak interactions~\cite{AxialFF}.
The nucleon axial form factors can be measured 
in quasi-elastic neutrino/antineutrino scattering 
with proton targets, by charged pion electroproduction on nucleons 
and also in the process of muon capture 
by protons~\cite{Bernard02,Schindler07a,Gorringe05}.
The value of the axial form factor at $Q^2=0$
is determined with great accuracy by neutron 
$\beta$ decay~\cite{Bernard02,PDG2014}.

The nucleon axial form factor has been calculated 
using different frameworks~\cite{AxialFF,NSTAR,Gaillard84,JDiaz04,Adamuscin08,Thomas84,Tsushima88,Shanahan13,Schlumpf93a,Boffi02,Merten02,BCano03,Silva05,Pasquini07,Anikin16a,Dahiya14a,Liu15a,Liu16a,Eichmann12,Chang13,Yamanaka14,Mamedov16a}.
Recently, also lattice QCD simulations 
of the nucleon axial form factors  
became available for several pion masses ($m_\pi$),
in the range $m_\pi =0.2$--0.6 GeV~\cite{Sasaki03,Edwards06,Yamazaki08,Bhattacharya14,Sasaki08,Yamazaki09,Bratt10,Alexandrou13,Alexandrou11a,Horsley14,ARehim15,Bali15,Bhattacharya16,Green17,Capitani17,Liang16,Berkowitz17,Yao17}.

In the present work our goal is to study the role 
of the valence quarks 
(leading twist approximation) and the role 
of the meson cloud ($\tau=5$) 
in the nucleon axial form factor $G_A$.
We consider in particular the holographic model from Ref.~\cite{Gutsche12a}, 
neglecting the gluon effects.
We assume that the gluon effects are included effectively 
in the quark structure through the gluon dressing.
In that case the next leading order correction 
is associated with the quark-antiquark excitations 
of the three valence quark core.
In this context the bare and the meson cloud contribution 
to the nucleon axial form factor are both expressed in 
terms of two independent parameters: $g_A^0$ and $\eta_A$,
associated with the quark axial and 
quark induced pseudoscalar couplings~\cite{Gutsche12a}.

To calculate the contributions associated 
with the nucleon bare core and the meson cloud 
we use the available experimental data and 
the results from lattice QCD, 
which help to constrain the contributions 
from the pure valence quark degrees of freedom,
and therefore fix also the contributions of the meson cloud component.
In the lattice QCD simulations with large pion masses
the meson cloud effects are very small, 
and the physics associated with the valence quarks can be better calibrated.

The results from lattice QCD cannot be directly related 
to the valence quark contributions to the axial form factor,
because the lattice calculations are not performed 
at the physical limit (physical quark masses).
The results from lattice can, however,
be extrapolated to the physical case with the
assistance of quark models that include a dynamic dependence on the quark mass.

Once fixed the parameters of the holographic model
by the empirical and lattice QCD data,
the holographic model can be used to estimate 
the fraction of the meson cloud contribution 
to the nucleon axial form factor.
This estimate can be compared to other 
estimates from quark models with meson cloud dressing.

We conclude at the end that the holographic model considered 
in the present work describes accurately  
the experimental data for the nucleon axial form factor,
and that the lattice QCD data with small pion masses
can be well approximated by the estimate    
of the valence quark contributions, in all ranges of $Q^2$.
We also conclude that the meson cloud contribution falls off very slowly 
with the square momentum transfer $Q^2$, 
much slower than estimates based on quark models.

This article is organized as follows.
In Sec.~\ref{secBackground}, we discuss the formalism  
associated with the study of the axial structure 
of the nucleon, including the axial current, parametrizations of the data,
results from lattice QCD,
as well as theoretical models based on a valence quark core with  
meson cloud dressing.
In Sec.~\ref{secHolography}, we present 
the holographic model for nucleon axial  form factor
considered in the present work.
The numerical results of the nucleon axial  form factor
and for the estimate of the meson cloud contributions
based on the holographic model appear in Sec.~\ref{secMesonCloud}.
The outlook and the conclusions are presented in Sec.~\ref{secConclusions}.

\section{Background}
\label{secBackground}

We now discuss the background associated 
with the study of the nucleon axial form factor.
We start with the representation of the axial current 
and the definition of the axial form factors.
Next, we summarize the experimental status
of the nucleon axial form factor $G_A$.
Later, we explain how the experimental data 
can be described within a quark model for the bare core,
combined with a meson cloud dressing of the core.
Finally, we discuss the results from lattice QCD
and how those results can be related with 
the function $G_A$ in the physical limit.

\subsection{Axial current}
\label{secCurrent}

The weak-axial transition between two nucleon states
with initial momentum $p$, final momentum $p'$, 
and transition momentum $q=p' -p$, is characterized  
by the weak-axial current~\cite{Bernard02,Schindler07a}
\ba
\left(J_5^\mu
\right)_a =
\bar u(p') \left[
G_A(Q^2) \gamma^\mu + G_P(Q^2) \frac{q^\mu}{2M} 
\right]  \gamma_5 u(p) \frac{\tau_a}{2},
\nonumber \\
\label{eqJA}
\ea  
where $M$ is the nucleon mass, $Q^2= -q^2$, 
$\tau_a$ ($a=1,2,3$) are Pauli isospin operators
and $u(p)$, $u(p')$ are the Dirac spinors associated 
with the initial and final states, respectively.
The functions $G_A$ and $G_P$ define, respectively, 
the axial-vector and the induced pseudoscalar form factors.
In the present work we restrict the analysis 
to the axial-vector form factor, refereed to  hereafter, 
simply as the axial form factor.
The leading order contribution for $G_P$ 
can be estimated considering the meson pole contribution, %which related $G_P$ with $G_A$
$G_P= \frac{4M^2}{m_\pi^2 + Q^2} G_A$,
derived from the partial conservation 
of the axial current~\cite{AxialFF,Bernard02,Thomas84,Sasaki08,Schindler07a,Gorringe05}.

Using the spherical representation ($a=0,\pm$)
we can interpret  $(J_5^\mu)_0$  as the current associated 
with the neutral transitions, $p \to p$ and $n \to n$ ($Z^0$ production),
and the current associated with $a=\pm$ 
with the $W^\pm$ production ($n \to p$ and $p \to n$ transitions).

\subsection{Experimental status}
\label{secPhenomenology}

The function $G_A$ can be measured by 
neutrino scattering and pion electroproduction off nucleons.
Both experiments suggest a dipole dependence 
$G_A(Q^2)= G_A(0)/(1+ Q^2/M_A^2)^2$, 
where the values of $M_A$ vary between 1.03 and 1.07 GeV
depending on the method~\cite{Bernard02,Schindler07a}.

To represent the experimental data in a general form 
we consider the interval between the two functions, $G_A^{{\rm exp}-}$
and $G_A^{{\rm exp}+}$, given by~\cite{AxialFF}
\ba
G_A^{{\rm exp}\pm}(Q^2)= \frac{G_A^0 (1 \pm \delta)}{
\left( 1 + \frac{Q^2}{M_{A\pm}^2} \right)^2},
\label{eqGAexp}
\ea
where $G_A^0=1.2723$ is the experimental value of $G_A(0)$~\cite{PDG2014},
$\delta =0.03$ is a parameter that expresses the precision 
of the data, and $M_{A-}=1.0$ GeV and  $M_{A+}=1.1$ GeV are,
respectively, the lower and upper limits from $M_A$
extracted experimentally.
The central value of the parametrization (\ref{eqGAexp}) 
can be approximated by a dipole with $M_A \simeq 1.05$ GeV.

Most of the data analysis are restricted to the region 
$Q^2 < 1$ GeV$^2$~\cite{Bernard02}.
The range of the variation associated with the 
parametrization of $G_A$ represented by Eq.~(\ref{eqGAexp}) 
is shown in Fig.~\ref{figMod0} by the red band.
The short-dashed-line represents the central value 
of the parametrization.

Recently the nucleon axial form factor 
was determined in the range $Q^2=2$--4 GeV$^2$
at CLAS/Jlab~\cite{Park12}.
The new data are consistent with the parametrization (\ref{eqGAexp}).

We discuss next, how the axial form factor 
can be estimated in the context of a quark model 
with meson cloud dressing of the valence quark core.

\subsection{Theory} 
\label{secTheory}

In a quark model with meson cloud dressing 
we can represent the physical nucleon  state 
in the form~\cite{AxialFF} 
\ba
\left|N \right>
= \sqrt{Z_N} \left[ \left| 3q\right> 
+ b_N \left| {\rm MC}\right> \right],  
\label{eqNucleonS}
\ea
where $\left| 3q\right>$ is the three-quark state 
and $b_N\left| {\rm MC}\right>$ is the meson cloud state.
The coefficient $b_N$ is determined by the normalization 
$Z_N (1+ b_N^2)=1$,
assuming that  $\left| {\rm MC}\right>$ is  normalized.

In this representation $Z_N= \sqrt{Z_N} \sqrt{Z_N}$
{\it measures} the probability of 
finding the $qqq$ state in the physical nucleon  state.
Consequently, $1-Z_N$ {\it measures} 
the probability of the meson cloud component in the physical nucleon state.

In Eq.~(\ref{eqNucleonS}), we include   
only the first correction for the meson cloud,
associated with the baryon-meson states.
In principle, we should also include corrections associated 
with baryon-meson-meson states.
In the case of the nucleon, however,
where the meson cloud is dominated by 
the pion cloud, the correction of the state 
$\left|N \pi \right>$ provides a good approximation 
to the physical nucleon  state.
In the case of $1- Z_N \simeq 0.3$ 
the correction associated with the two-pion correction 
is attenuated by the factor $(1- Z_N)^2 \simeq 0.09$.

In the calculation of the axial form factors,
in order to take into account the contribution
of the meson cloud in the form factors 
at the physical limit, 
one needs to correct the function $G_A^{\rm B}$ by the factor $Z_N$, 
which quantifies the contribution of the bare core to $G_A$~\cite{AxialFF}.
The effective contribution from $G_A^{\rm B}$ to the 
physical $G_A$ becomes then $Z_N G_A^{\rm B}$.
More generically, we can write 
\ba
G_A = Z_N G_A^{\rm B} + (1-Z_N) G_A^{\rm MC},
\label{eqGAsp}
\ea
where the second term accounts for the contribution 
from the meson cloud.
The function $G_A^{\rm MC}$ is the unnormalized 
meson cloud contribution,
estimated when we drop the valence quark contribution.

Hereafter, we use the expression bare contribution 
to refer the first term of Eq.~(\ref{eqGAsp})
and meson cloud contribution 
to refer the second term of Eq.~(\ref{eqGAsp}).

An alternative representation of the meson cloud term  
is $(1- Z_N) G_A^{\rm MC} = Z_N \tilde G_A^{\rm MC}$~\cite{AxialFF}.
To convert to $G_A^{\rm MC}$, one uses 
$G_A^{\rm MC} = \tilde G_A^{\rm MC}/(1/Z_N -1)$.
The function $\tilde G_A^{\rm MC}$ can be extracted from the data,
as discussed in Ref.~\cite{AxialFF}.

\subsection{Information from lattice QCD}
\label{secLattice}
  
Another source of information about the axial 
structure of the nucleon are the lattice QCD simulations.
In lattice QCD, one can simulate the dynamic of QCD in a discrete space-time.
Since simulations with very small grids and large volumes 
are very costly, most of the simulations are performed
for large values of the pion mass,
and the obtained results correspond to quark masses 
larger than the physical quark masses.
For those reasons, some care is necessary in the interpretation 
of the lattice QCD results, and in the extrapolations,
to the continuous limit, to the infinite volume limit, 
and to the physical limit (physical masses)~\cite{Bali15,Bhattacharya16}. 

Nevertheless, lattice QCD can be used 
to make a connection with results from quark models.
In those conditions, lattice QCD can help us 
to understand the role of the valence quarks in the structure form factors.
Since in lattice QCD simulations with large pion masses 
the effect of the meson cloud dressing is significantly reduced,
those simulations can be used to estimate 
the contribution of the form factors that 
are the direct consequence of the valence quark effects.
Contrary to the lattice QCD calculations 
of the electromagnetic form factors,
the nucleon axial form factor, due to its 
isovector character, has no contributions 
associated with the disconnected diagrams
in the continuous limit~\cite{AxialFF,Alexandrou11a,Alexandrou13,ARehim15},
and can therefore be directly compared to the experimental data.

The axial form factor and  
the induced pseudoscalar form factor
have been calculated in lattice QCD simulations 
for several values of the pion mass 
at $Q^2=0$~\cite{Sasaki03,Edwards06,Yamazaki08,Bhattacharya14,Liang16}, 
and for finite $Q^2$~\cite{Sasaki08,Yamazaki09,Bratt10,Alexandrou13,Alexandrou11a,Green17}.
Simulations with large volumes and pion masses 
in the range 0.25--0.5 GeV suggest that the values of $G_A$
near $Q^2=0$ are  generally restricted to $G_A(0)=1.1$--1.2.
Those results indicate that based only on
the contributions of the valence quarks it is not possible to reach 
the experimental value  $G_A(0) \simeq 1.27$.
This underestimation can be inferred as a sign 
that the meson cloud contribution to $G_A$ is positive.

\begin{figure}[t]
\vspace{.5cm}
\centerline{
\mbox{
\includegraphics[width=3.0in]{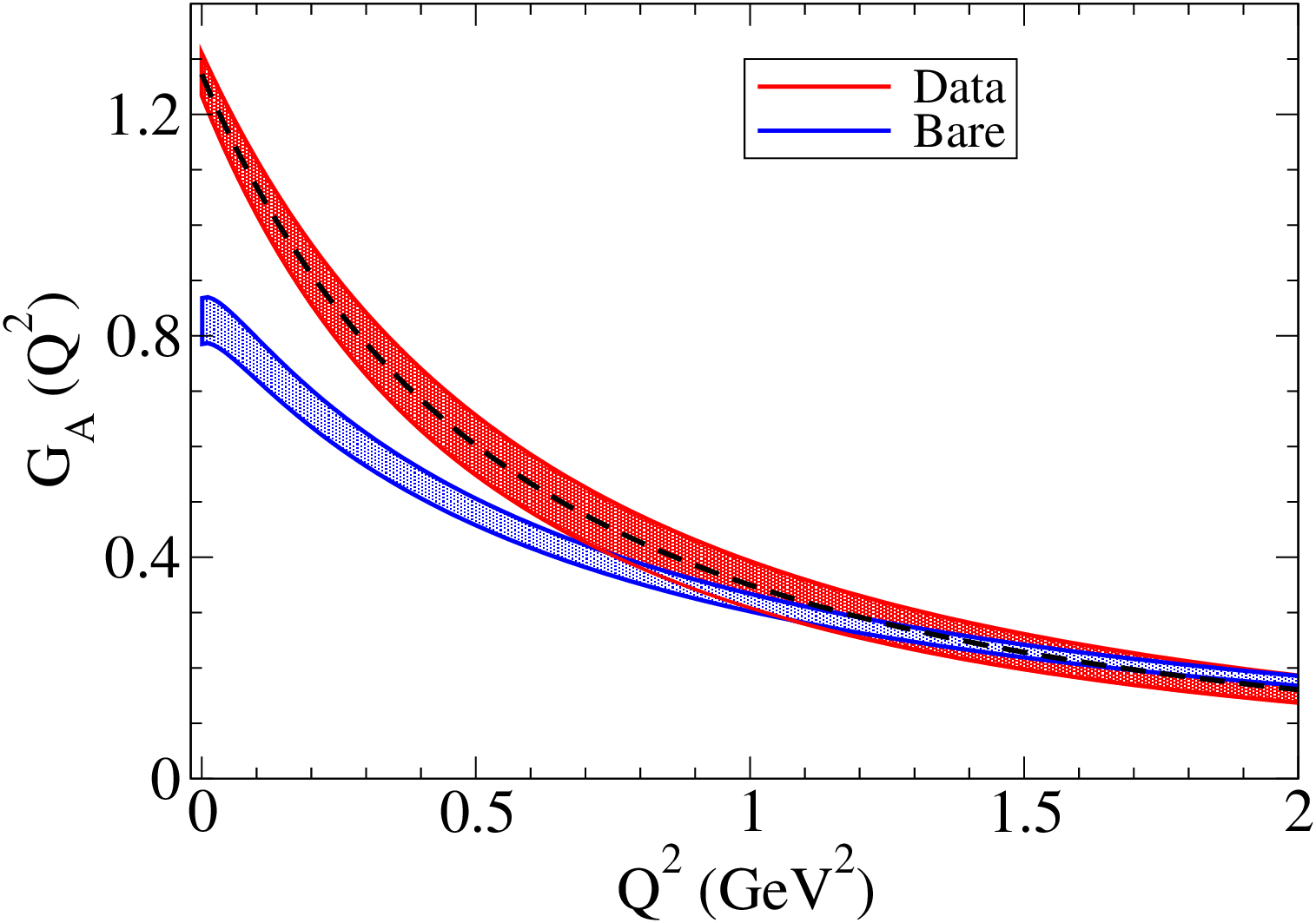} }}
\caption{\footnotesize{
Experimental parametrization of the data $G_A^{\rm exp}$ 
according to Eq.~(\ref{eqGAexp}), at red, 
combined with the estimate of the contribution $Z_N G_A^{\rm B}$
extracted from the lattice QCD data, at blue.
The short-dashed-line indicates the central value 
of Eq.~(\ref{eqGAexp}).}}
%\vspace{-1cm}
\label{figMod0}
\end{figure}

Estimates of $G_A(0)$ near the physical point 
can be found in Refs.~\cite{Horsley14,ARehim15,Bali15,Bhattacharya16,Capitani17,Berkowitz17,Yao17}.
Lattice QCD simulations with smaller pion masses 
may include some meson cloud effects 
and may also be affected by significant 
finite volume effects, which tend to underestimate 
the value of $G_A(0)$ compared to 
the infinite volume limit~\cite{Yamazaki08,Yamazaki09}.

The study of the valence quark effects in the nucleon axial form factor 
can also be performed considering a constituent quark model where  
the parameters associated with the properties of the quarks  
are adjusted in order to describe the results from lattice QCD.
In this case the decisive parameter is the variable that regulates 
the quark mass which can be converted into the mass of the pion   
associated with the lattice QCD regime.

One can then extrapolate the valence quark 
contribution of $G_A$ in the physical limit from the lattice QCD results,  
using a quark model,  
if the parameters of the model are defined in terms of the pion mass.
It is worth noticing, however, that the function $G_A$ 
extrapolated to the case $m_\pi \to m_\pi^{\rm phys}$ 
($ m_\pi^{\rm phys}$ represent the physical pion mass),
which may be interpreted as $G_A^{\rm B}$ (bare contribution), 
does not represent in fact the bare contribution to the physical form factor.
This happens, because in the physical limit, 
one needs to take into account the effect 
of the meson cloud dressing and its impact
in the physical nucleon  wave function, as shown in Eq.~(\ref{eqGAsp}).
The effective contribution to the physical $G_A$ is then $Z_N G_A^{\rm B}$, 
where $G_A^{\rm B}$ is the contribution from the valence quark component,
estimated from lattice QCD, and extrapolated to the physical case.
An example of a quark model with proprieties mentioned 
above is the model from Ref.~\cite{AxialFF}.

In Ref.~\cite{AxialFF}, the covariant spectator quark model 
is applied to the study of the axial structure of 
the nucleon in the lattice QCD regime, and in the physical regime.
In the covariant spectator quark model, 
hereinafter referred to simply as the spectator model,
the nucleon is described as a three valence quark system 
and the radial wave functions are expressed in terms of momentum scale parameters 
determined in the study of the nucleon electromagnetic structure~\cite{Nucleon}.
The nucleon valence quark wave function is represented  
by a mixture of two states: the dominant $S$-wave and a small $P$-wave, 
as in other quark models~\cite{AxialFF,Thomas84}.
The quark substructure is parametrized 
by quark electromagnetic and axial form factors, 
which simulate effectively the internal structure of the constituent quarks,
resulting from the interactions with quark-antiquark pairs
and from the quark-gluon dressing~\cite{Nucleon,Omega1}.
The parameters of the spectator model 
associated with the valence quark structure are 
first fixed by the lattice QCD data and 
the results are later extended to the physical limit.

We can summarize the method used in 
Ref.~\cite{AxialFF} by the following steps:
\begin{itemize}
\item
Calibration of the parameters associated with the 
valence quark structure 
(quark form factors and fraction of $P$-state mixture) 
using lattice QCD data.
\item
Extend the result of $G_A(Q^2,m_\pi)$ 
to the physical limit ($m_\pi \to m_\pi^{\rm phys}$)
defining the function $G_A^{\rm B}(Q^2)$.
\item
Use experimental data to determine the factor $Z_N$ 
associated with the normalization of the physical nucleon state, 
according to
\ba
G_A^{\rm exp} (Q^2)\simeq Z_N G_A^{\rm B}(Q^2),
\ea
in the region $Q^2 > 1$ GeV$^2$, where the meson  cloud effects are expected to be small.
This procedure establishes the proportion of meson cloud 
in the physical nucleon  state.
\item
The contribution from the meson cloud to $G_A$ 
can then be estimated by the difference:
$G_A^{\rm exp}- Z_N G_A^{\rm B}$ 
for small $Q^2$ ($Q^2 < 1$ GeV$^2$).
\end{itemize}

The connection between the spectator model and the lattice regime 
is performed using wave functions 
dependent of the mass of the nucleon 
(physical mass replaced by lattice mass),
and quark form factors parametrized in terms 
of the vector dominance mechanism~\cite{AxialFF}.
In the lattice QCD regime, the vector meson physical masses 
are replaced by the masses of vector mesons in lattice.
Except for the masses (baryons and vector mesons) all the parameters 
of the wave functions and quark form factors are determined 
by fits to the lattice QCD data.
Check Ref.~\cite{AxialFF} for more details about 
the parametrization of the quark axial structure.
More details about the extension of the spectator model to 
the lattice QCD regime can be found in  Refs.~\cite{Lattice,LatticeD,NucleonMC3,Octet2,Omega1,Omega2}.

The function $G_A^{\rm B}$ extrapolated from lattice QCD
using the spectator model based on the previous procedure
is presented in Fig.~\ref{figMod0}, by the blue band.
The $P$-state mixture is 25\%~\cite{AxialFF}.
The accuracy of the parametrization for $G_A^{\rm B}$ 
is then limited by the precision of the lattice data.
Since the lattice data can be very accurate for small $Q^2$
($\sim 1\%$) and have large errorbars for $Q^2=2$--4 GeV$^2$
($\sim 10\%$), we consider an average error of 5\%.

For future reference, we mention that the parametrization 
of the meson cloud contribution in the spectator model 
can be represented by~\cite{AxialFF}
\ba
G_A^{\rm MC}(Q^2) =
\frac{G_A^{{\rm MC}0}}{\left(1 + \frac{Q^2}{\Lambda^2} \right)^4}, 
\label{eqGMCspec}
\ea
where $G_A^{{\rm MC}0} =1.68$ and $\Lambda=1.05$ GeV. 
Here,  $\Lambda$ is the average of the two 
cutoffs used in the parametrization (\ref{eqGAexp}).
We recall that the effective contribution 
of $G_A^{\rm MC}$ to $G_A$ is the result of 
the product $(1-Z_N)G_A^{\rm MC}$.

\section{Holographic Model}
\label{secHolography}

Different holographic models 
have been applied to the systems ruled by QCD. 
Those models can be classified into two main categories:
the bottom-up approach and  the top-down approach.
The top-down approach is related to supersymmetric strings 
and it has the base of the $D$-brane 
physics~\cite{Nawa07,Hong07,Hong08,Sakai05,Hashimoto08}.
The bottom-up approach is more phenomenological 
and derive the QCD proprieties  in the confining regime
using 5D-fields in AdS space~\cite{Brodsky15,Brodsky08a,Teramond09a,Karch06,Gutsche12a,Gutsche13a}.

In the present work we consider a  bottom-up
approach where the confinement is included through 
a potential $U_F(z)$ (soft-wall approximation).
We consider in particular 
the holographic soft-wall model from Ref.~\cite{Gutsche12a}
for the nucleon axial form factor.

In holographic QCD the particle fields
$\Psi$ and the source fields (electromagnetic and axial) 
are represented in terms of the coordinates $(x,z)$,
where $x$ belongs to the usual 4D space 
and $z$ is the holographic variable.
To describe the structure of the baryons 
we define fermion fields $\Psi(x,z)$, 
which encode the proprieties of the  baryons.
Those fermion fields can be decomposed 
into different modes $\Psi_n$ ($n=0,1,2,...$)
which are the holographic analogous 
of the baryon wave functions~\cite{Gutsche12a,Gutsche13a,Abidin09}.

For the description of the nucleon structure 
we start by constructing the fermion fields $\Psi_\pm (x,z)$
associated with the spin $J=1/2$, where $\pm$ are 
the left- and right-handed ($L/R$) components 
of the nucleon radial excitations doublets. 
The axial structure is introduced by 
the 5D axial field $\hat {\cal A}_i(x,z)$,
where $i=\pm$.
Following Ref.~\cite{Gutsche12a}, we represent 
the axial structure in the form 
\ba
\hat {\cal A}_i(x,z)= 
 \hat {\cal A}_i^{(1)}(x,z) + \hat {\cal A}_i^{(2)}(x,z) 
+ \hat {\cal A}_i^{(3)}(x,z).
\label{eqAxial}
\ea
The different terms describe the possible structures 
associated with the axial interaction in 5D.

The first term is the minimal axial-vector coupling 
\ba
 \hat {\cal A}_i^{(1)}(x,z)= g_A^0 \Gamma^M \gamma_5 A_M(x,z) \frac{\tau_3}{2}, 
\ea
where $\Gamma^M$ ($M=0,1,2,3,z$) is the 5D gamma matrix,
$A_M(x,z)$ is the holographic analogous of the axial field
and $\tau_3= \mbox{diag}(1,-1)$ is the Pauli isospin matrix.
The function $A_M(x,z)$ is constrained by the gauge condition
$A_z(x,y) = 0$~\cite{Gutsche12a}. 
The second term represents a nonminimal coupling,
the holographic analogous of the induced 
pseudoscalar coupling 
\ba 
\hat {\cal A}_i^{(2)}(x,z)= \eta_A 
\left[\Gamma^M, \Gamma^N \right] \gamma_5 A_{MN} (x,z) \frac{\tau_3}{2}, 
\ea
where $A_{MN} = \partial_M A_N- \partial_N A_M$.
The final term is an axial-type coupling 
proportional to the nucleon isovector charge
\ba
 \hat {\cal A}_i^{(3)} (x,z)= \mp \Gamma^M A_M(x,z) \frac{\tau_3}{2}. 
\ea

The fermion fields, mentioned above can be expressed in the 
Weyl representation in the form 
\ba
\Psi_{\pm,n} (x,z) = 
z^2\left(\begin{array}{c}
F_{L/R,n}(z)  \\
\pm F_{R/L,n}(z)  \\
\end{array} \right)
\chi_n (x),
\label{eqWF}
\ea
where $\chi_n(x)$ is a two-component spinor and 
the functions $F_{L/R,n}(z)$ are 
solutions of Schr\"{o}dinger-type  
wave equations in the variable $z$~\cite{Gutsche12a,Gutsche13a,RoperHol}.
For simplicity, we omitted the isospin indices.
The nucleon case corresponds to the first mode ($n=0$).
More details can be found in Refs.~\cite{Brodsky15,Gutsche12a,Gutsche13a,Abidin09}.

The axial transition current is calculated 
considering the overlap of the holographic nucleon  
fields associated with the initial and final states 
with the axial field (\ref{eqAxial}).
From the axial transition current 
we can extract the holographic expressions 
for the axial form factor $G_A$ 
according with the number of constituents.

\subsection{Axial form factor}

In Ref.~\cite{Gutsche12a}, the contributions associated 
with the first Fock states are studied in detail,
and the effects of the 3, 4 and 5 parton components 
are calculated explicitly.
Those contributions are associated, respectively, 
with the $qqq$ state (3-quark, index $\tau=3$),
the $(qqq)g$ state (3-quark-gluon,  index $\tau=4$),
and the $(qqq)\bar q q$ state (3-quark-quark-antiquark, index $\tau=5$).
Neglecting the contributions associated with the gluon states, 
we can write the nucleon axial form factor $G_A$ in the form 
\ba
G_A(Q^2) = c_3 G_A^{\rm B} (Q^2) + c_5 G_A^{\rm MC}(Q^2),
\label{eqGA0}
\ea
where $G_A^{\rm B}$ is the bare contribution associated 
with the $qqq$ state ($\tau=3$) and 
 $G_A^{\rm MC}$ is the meson contribution associated 
with the $(qqq)q \bar q$ state ($\tau=5$).
The coefficients $c_\tau$ specify the weight of 
the $\tau$-component of the Fock state, and 
are in the present approximation restricted to  $c_3 + c_5=1$.
According to Ref.~\cite{Gutsche12a}, 
the components $G_A^{\rm B}$ and $G_A^{\rm MC}$ 
can be represented 
in terms of $a= \frac{Q^2}{4 \kappa^2}$, as
\ba
G_A^{\rm B} (Q^2) & = & 
\left[ g_A^0 + \frac{a}{6} (g_A^0 -1) \right] G_1  \nonumber \\
& & + \frac{\eta_A}{12} a(2 a + 17) G_2, 
\label{eqGAB}\\
  G_A^{\rm MC} (Q^2) & = & 
\left[ g_A^0 + \frac{a}{10}(g_A^0 -1) \right] G_3  \nonumber \\
& & + \frac{\eta_A}{30} a(4 a + 49) G_4,
\label{eqGAMC}
\ea
where the functions $G_i$ ($i=1,2,3,4$) have the following form
\ba
G_1 &= & \frac{1}{\left(1 + a \right) \left(1 + \frac{a}{2} \right)
\left(1 + \frac{a}{3} \right)}, \\
G_2 &= & \frac{1}{\left(1 + a \right) \left(1 + \frac{a}{2} \right)
\left(1 + \frac{a}{3} \right) \left(1 + \frac{a}{4} \right) }, \\
G_3 &= & \frac{1}{\left(1 + a \right) \left(1 + \frac{a}{2} \right)
\left(1 + \frac{a}{3} \right) \left(1 + \frac{a}{4} \right) 
\left(1 + \frac{a}{5} \right) 
}, \\
G_4 &= & \frac{1}{\left(1 + a \right) \left(1 + \frac{a}{2} \right)
\left(1 + \frac{a}{3} \right) \left(1 + \frac{a}{4} \right) 
\left(1 + \frac{a}{5} \right) \left(1 + \frac{a}{6} \right) 
}. \nonumber \\
& &
\ea
Recall that in the previous equations
$\kappa$ is the holographic mass scale.
In the following, we consider the value $\kappa =0.385$ GeV, 
in order reproduce approximately the $\rho$ mass ($m_\rho \simeq 770$ MeV).
The holographic estimate of the nucleon 
mass is then $2 \sqrt{2}\kappa \simeq 1.09$ GeV,
a bit above the experimental value.

From Eqs.~(\ref{eqGAB})-(\ref{eqGAMC})
we can conclude that at large $Q^2$:
$G_A^{\rm B} \propto 1/Q^4$ and $G_A^{\rm MC} \propto 1/Q^8$.
As a consequence, the meson cloud contribution
falls off faster than  the bare contribution.
One can then expect that $G_A^{\rm MC}$ become negligible for values 
of $Q^2$ larger than a certain scale. 
One of the goals of the present study is to estimate that scale.

Concerning the decomposition of the bare and meson cloud 
contributions in terms of the pole structure of the functions $G_i$,
some discussion is in order.
The present representation in terms of the poles on $a$ 
is a direct consequence of the calculation of 
the axial form factors based on the axial coupling 
(\ref{eqAxial}) and the wave functions (\ref{eqWF}).
The present pole structure of the functions $G_i$
is expected for the calculation of the electromagnetic 
form factors~\cite{Gutsche12a,Gutsche13a,Abidin09},
and can be interpreted in terms 
of the vector meson dominance (VMD) 
mechanism~\cite{Maldacena99,Brodsky15,Grigoryan07a,Hong07,Hong08,Hashimoto08,Sakurai60,Bauer79,Gari84,Lomon01,Sakai05,Nawa07}.
It differs, however, from other approaches, 
which represent the axial form factors 
in terms of axial-vector meson poles~\cite{Adamuscin08,Hashimoto08,Hong07}.
Later on, we discuss parametrizations
based on the axial-vector meson masses.

\section{Estimations of the meson cloud contributions}
\label{secMesonCloud}

From the holographic parametrizations 
of the axial form factors~(\ref{eqGAB})-(\ref{eqGAMC}), 
one can conclude that at $Q^2=0$, 
the bare contribution is $c_3 g_A^0$,
and the meson cloud contribution is $(1- c_3) g_A^0$.
Adding the two terms, one obtains $G_A(0) = g_A^0$.

From the previous result, we conclude that in a holographic model, 
the description of the function $G_A$ near $Q^2=0$
may require contributions from the bare and from the meson cloud components.

Since both components, $G_A^{\rm B}$ and  $G_A^{\rm MC}$,
depend on the couplings $g_A^0$ and $\eta_A$, 
we may question if a global fit of the 
parameters $c_3$, $g_A^0$ and $\eta_A$ 
to the empirical parametrization of $G_A$ given by Eq.~(\ref{eqGA0})
is sufficient to fix the two components of $G_A$,
without any additional constraints from the physics associated 
with the bare core or with the meson cloud.

To test the previous hypothesis, we start 
performing a global fit of the parameters $c_3$, $g_A^0$ and $\eta_A$, 
to the  empirical parametrization of the data (\ref{eqGAexp}),
obtaining a naive estimation of the bare contribution.
Later on, we discuss if the calibration 
of the components $G_A^{\rm B}$ and $G_A^{\rm MC}$ may 
be improved using constraints associated with the function $G_A^{\rm B}$,
extracted from lattice QCD.

In the following, we consider several parametrizations 
of the data in the region $Q^2=0$--2 GeV$^2$.
In this region, we expect that both, bare and meson cloud components, 
have relevant contributions, although,
we expect also a significant
reduction of the meson cloud contribution for $Q^2 > 1$ GeV$^2$ (faster falloff).
We recall that most of the available data are in the region $Q^2 < 1$ GeV$^2$.

\subsection{Naive estimations of the bare contribution}

An unconstrained fit of the holographic model (\ref{eqGA0})
to the parametrization of the data (\ref{eqGAexp}),
results in an excellent description 
of the central value from $G_A^{\rm exp}$.
The parameters obtained from the fit 
are: $g_A^0 \simeq 1.2723$ (experimental value),
$\eta_A \simeq 0.45$ and $c_3 \simeq 1.45$.
The coefficient associated with the meson cloud term is 
then $c_5 \simeq -0.45$, which correspond to a negative contribution 
of the meson cloud component.
Since, from the lattice QCD studies,
we expect positive contributions to the meson cloud, 
we discard this solution as a physical solution.
It is worth noticing, however,
that this first fit provides a parametrization
very close to the model originally derived in Ref.~\cite{Gutsche12a},
where $\kappa =0.383$ GeV, $\eta_A = 0.5$ and $c_5= -0.41$.
In that model there is also a small  contribution from 
the $(qqq)g$ component with a weight $c_4= 0.16$. 

The result of the fit is indistinguishable from the 
central value from (\ref{eqGAexp}) represented in Fig.~\ref{figMod0},
by the short-dashed-line.
Recall that the red band represents  
the limits of the experimental parametrization.

In order to constrain the holographic model 
to positive contributions for the meson cloud, 
we refit the function (\ref{eqGA0}) to the data 
under the condition $c_5 > 0$, which is 
equivalent to $c_3 < 1$.
The result of the this fit is a solution with $c_3 \simeq 1$
combined with $g_A^0 \simeq 1.27$ (experimental value) and 
$\eta_A \simeq 0.68$.
Since $c_5 = 1-c_3 \simeq 0$, this solution 
corresponds to the case $G_A(Q^2) \equiv G_A^{\rm B}(Q^2)$.
Also, this solution is at the top of 
the empirical parametrization (\ref{eqGAexp}),
and it cannot be distinguished from the previous parametrization
(see Fig.~\ref{figMod0}).

One then concludes, that without 
additional constraints relative to 
the magnitude of the bare contribution (or meson cloud), 
a holographic model with no meson cloud contribution 
describes well the empirical data for $G_A$.

Another important conclusion is that the  
 experimental parametrization (\ref{eqGAexp}) (central value)
can be reproduced by a combination of the functions $G_i$ 
associated with the poles $4(n+1)\kappa^2$ ($n=0,1,..,5$).
Thus, below 2 GeV$^2$, the holographic model 
is numerically equivalent to a dipole parametrization,
whether we include the meson cloud or not,
as discussed above.

\begin{table}[t]
\begin{center}
\begin{tabular}{c c c c}
\hline
\hline
$g_A^0$ & $\eta_A$ & $c_3$ & $\chi^2(G_A^{\rm B})$  \\
\hline
1.273  &  1.072\spQ &  0.702 & 2.54 \\
{\bf 
1.200} &  1.083\spQ &  0.721 & 2.06  \\
1.125  &  1.094\spQ &  0.743 & 1.65 \\
\hline
\hline
\end{tabular}
\end{center}
\caption{Parameters of the models and respective value 
of chi-square per data point associated with $G_A^{\rm B}$.}
\label{tableMods}
\end{table}

\subsection{Using lattice QCD information}

A more qualified
description of the axial form factor can be obtained
if we use the information relative to the function $G_A^{\rm B}$, 
extracted from the study of the lattice QCD data.

As discussed in Sec.~\ref{secTheory}, 
the function $G_A^{\rm B}$ does not represent the effective 
contribution of the quark core to the form factor $G_A$,
because the effect of the meson cloud component 
in the physical  nucleon  state 
needs to be taken into account.
As a consequence only $Z_N G_A^{\rm B}$ contributes to 
the physical form factor $G_A$, where  
$Z_N$ gives the probability associated with the $qqq$ component
in the physical nucleon, according to Eq.~(\ref{eqGAsp}).

One can now correlate the holographic relation~(\ref{eqGA0}),
with the expression for $G_A$ derived 
from a valence quark model with meson cloud dressing~(\ref{eqGAsp}),
identifying $c_3 \equiv Z_N$.
Note, however, that this relation is valid  
only when $c_3 \le 1$, because $Z_N$ is by definition limited to 
\mbox{$Z_N \le 1$.}
The upper limit represents the valence quark limit,
when there is no meson cloud (the coefficient of the meson cloud term is $1-Z_N =0$).

To take into account the information relative to the bare component,
we include in the fit the function $G_A^{\rm B}$
extrapolated from lattice QCD, with the assistance of the spectator model,  
as discussed in Sec.~\ref{secLattice}.
In the numerical fits of $G_A^{\rm B}$, we consider 
41 datapoints in the region $Q^2=0$--2 GeV$^2$.

A new class of  parametrizations is then 
obtained when we adjust the parameters $c_3$,
$g_A^0$ and $\eta_A$ to the parametrizations $G_A^{\rm B}$ 
(extracted from lattice) and $G_A^{\rm exp}$.
As we show next, the description of $G_A$ depends crucially on $g_A^0$.

\begin{figure}[t]
\vspace{.5cm}
\centerline{
\mbox{
\includegraphics[width=3.0in]{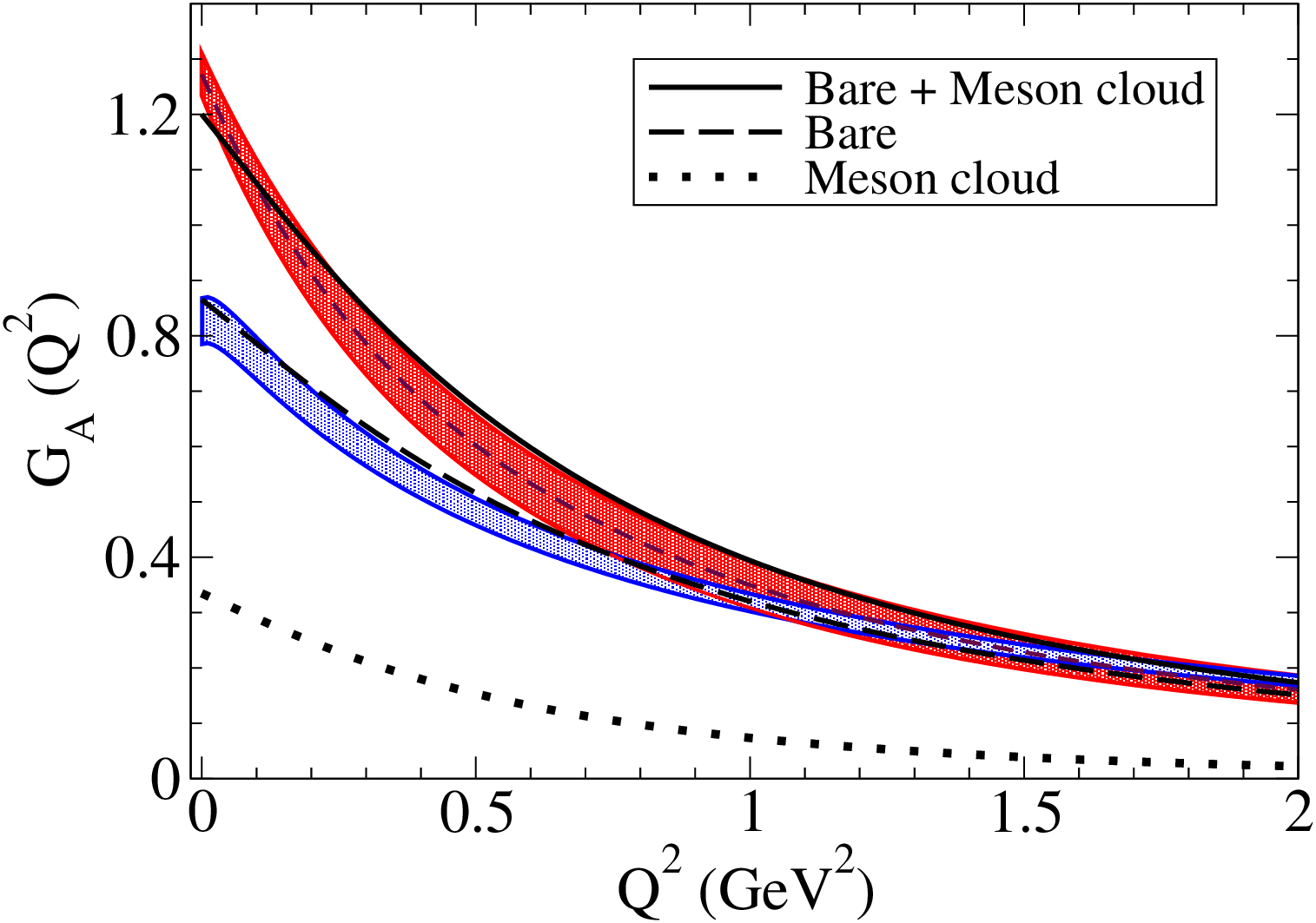} }}
\caption{\footnotesize{
Results of the fit of the axial form factor $G_A$ using $g_A^0=1.2$.
The solid-line represent the function $G_A$, 
the dashed-line represent the bare contribution, $c_3\, G_A^{\rm B}$,
and the dotted-line meson cloud contribution, $c_5\, G_M^{\rm MC}$.
The red and blue bands have the same meaning of Fig.~\ref{figMod0}.}}
%\vspace{-1cm}
\label{figMod2}
\end{figure}

We consider three fits to the functions $G_A^{\rm exp}$ and $G_A^{\rm B}$.
First, we consider a free fit using the parametrizations described below,
which fails to describe the low $Q^2$ region of $G_A$.
In a second fit, we attempt to describe in 
more detail $G_A^{\rm exp}$ near $Q^2=0$, imposing $G_A(0) = 1.2723$,
but overestimate the $G_A^{\rm B}$  parametrization.
Finally, we consider an intermediate fit which   
compromises the description of the functions  $G_A^{\rm exp}$ and $G_A^{\rm B}$.

In a global fit with no constraints in $g_A^0$
using the empirical parametrization (\ref{eqGAexp}),
we obtain $g_A^0= 1.125$ 
[error of 3\% for $G_A(0)$].
The  remaining parameters are presented in the last row of Table~\ref{tableMods}.
Note, in particular, that in this fit the contribution 
of the meson cloud in the physical nucleon state is 26\%
($c_5= 1-c_3 \simeq 0.26$),  
and that the fit to the function $G_A^{\rm B}$ has 
a chi-square per datapoint of 1.65.
Since the parametrization gives  $G_A(0) = g_A^0 = 1.125$, 
we can conclude that this fit underestimates the experimental data 
near $Q^2=0$ (the experimental value is 1.2723).

To improve the description of $G_A$ near $Q^2=0$,
one needs to constrain the values of $g_A^0$ 
to values closer to the experimental value for $G_A(0)$.
This can be done varying the values of $\eta_A$ and $c_3$, 
and keeping $g_A^0 =1.2723$. 
In this case, one obtains $c_3=0.702$,
(first row in Table~\ref{tableMods}),
but decreases the quality  
of the description of the component $G_A^{\rm B}$
(chi-square per data point of 2.54).

Finally, we consider a parametrization  with an intermediate $g_A^0$, using $g_A^0 =1.2$.
In this case we also obtain a description 
of $G_A^{\rm exp}$ closer to the range of one standard deviation 
and  also a fair description 
of the function $G_A^{\rm B}$ (chi-square per datapoint of 2.06).
The contribution of the meson cloud in 
the physical nucleon is in this case 28\%
(second row in Table~\ref{tableMods}).

The graphical representation of the last 
parametrization is presented in Fig.~\ref{figMod2}.
The function $G_A$ is represented by the solid-line;
the function $c_3 G_A^{\rm B}$ is represented by the dashed-line,
and the meson cloud contribution $c_5 G_A^{\rm MC}$ 
is represented by the dotted-line.

From  Fig.~\ref{figMod2}, one can conclude that 
the fit associated with $g_A^0 =1.2$
provides the simultaneous description 
of the parametrizations $G_A^{\rm exp}$ and $G_A^{\rm B}$
(red and blue bands, respectively),
since the lines associated with $G_A$ (solid-line)
and $G_A^{\rm B}$ (dotted-line) are almost always inside 
the respective bands (one standard deviation).
One can also see, that the meson cloud contribution 
(dotted-line) falls off faster than the bare contribution (dashed-line).
This falloff is discussed in more detail in the following sections.

The parametrization from  Fig.~\ref{figMod2}
corresponds to a meson cloud admixture coefficient $c_3= Z_N =0.72$, 
meaning that the 
meson cloud component accounts for 28\% 
of the physical nucleon state, as mentioned above.
This estimate is very close to the estimates 
from the spectator model from Ref.~\cite{AxialFF} (27\%)
and also from the Cloudy Bag Model from Ref.~\cite{Shanahan13} (29\%).
The estimates from the perturbative chiral quark model~\cite{Liu15a,Liu16a}
are also similar to our results for 
the bare and meson cloud contributions 
to the nucleon axial form factors, at low $Q^2$.

We can then conclude that our estimate  of the amount of the meson cloud 
is close to other estimates of the that effect (around 30\%).

\subsection{Discussion about $g_A^0$}

We now discuss in more detail the effect of the parameter $g_A^0$ in the calculations.
As mentioned previously, the holographic results for $G_A(Q^2)$ 
are strongly dependent on $g_A^0$.
Large values of $g_A^0$ ($g_A^0 > 1.3$) overestimate the low $Q^2$ data.
Small values of  $g_A^0$ ($g_A^0 < 1.1$) underestimate the low $Q^2$ data.
The constraints from lattice QCD favors values of $g_A^0$ smaller than 1.27.
Recall that the results 
obtained in lattice QCD simulations for $G_A(0)$  
are in general restricted to $G_A (0) = 1.1$--1.2.

In order to check the range of $g_A^0$ preferred by the lattice data, 
we start by comparing the holographic models directly 
with the lattice QCD data.
Notice that the holographic model includes 
a bare and a meson cloud component.
Since the $Q^2$-dependence of the lattice data 
varies with the pion mass, we select lattice QCD data 
associated with the pion masses not to far way from the physical limit.
We consider in particular data associated with $m_\pi= 213$, 
260 and 262 MeV from Refs.~\cite{Alexandrou13,Alexandrou11a}.

\begin{figure}[t]
\vspace{.5cm}
\centerline{
\mbox{
\includegraphics[width=3.0in]{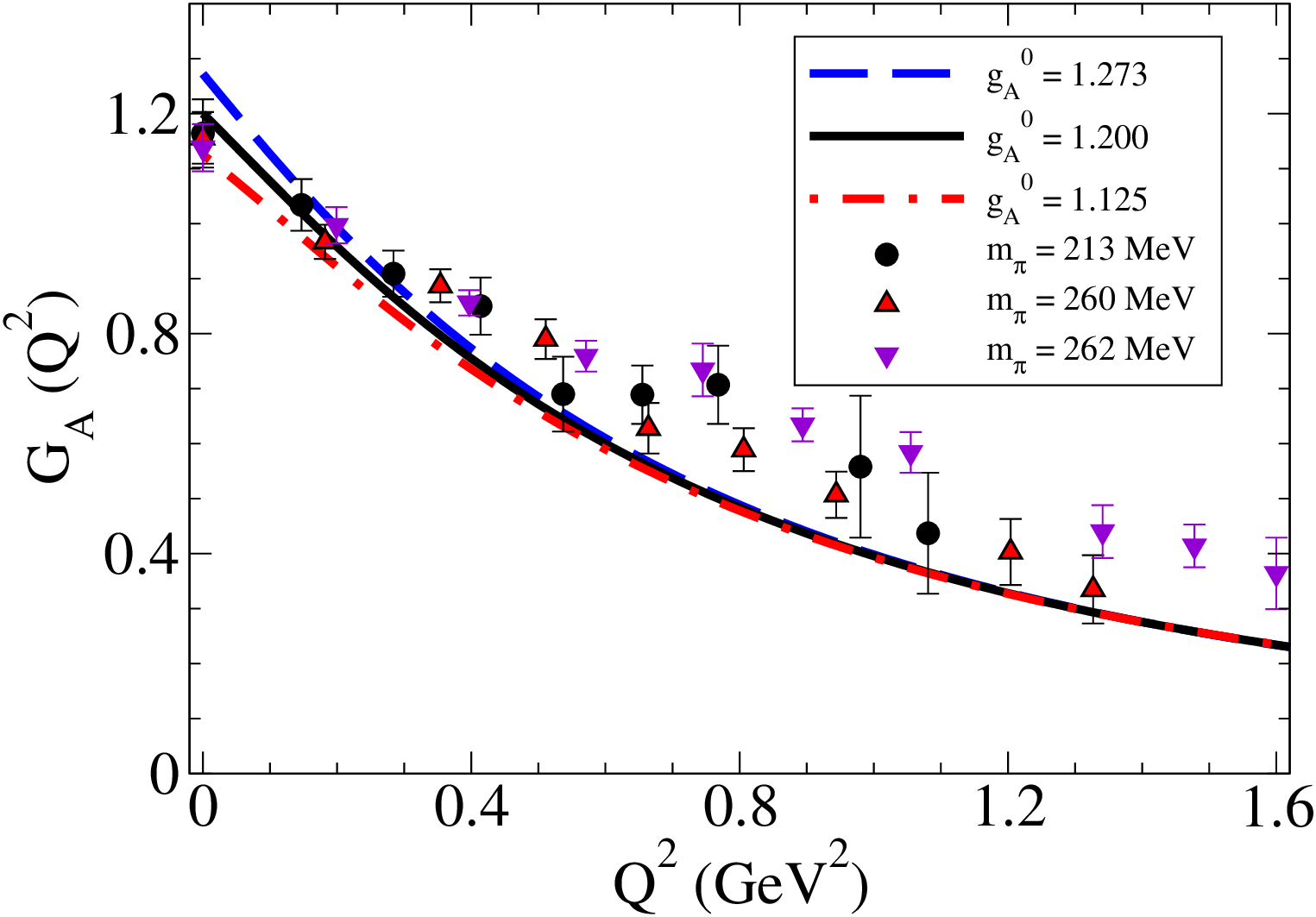} }}
\caption{\footnotesize{
Comparison of the holographic models (full result) with the
lattice QCD data~\cite{Alexandrou13,Alexandrou11a}.
The models are labeled with the value of 
$g_A^0$ presented in Table~\ref{tableMods}.
}}
%\vspace{-1cm}
\label{figLat2}
\end{figure}

The comparison with the lattice QCD data is presented 
in Fig.~\ref{figLat2} for the parametrizations 
from  Table~\ref{tableMods}, labeled by $g_A^0=1.125$, 1.2 and 1.273. 
In the figure we can observe a good agreement 
with the data for the parametrizations with $g_A^0=1.125$ and 1.2
below $Q^2=0.3$ GeV$^2$, and a systematic deviation for larger 
values of $Q^2$ for all the parametrizations.

There are in principle two main reasons for the deviation 
between the lattice data and the holographic estimates.
On one hand the holographic model under discussion 
is developed for the physical limit.
Therefore the bare and the meson cloud components 
are estimates for $m_\pi= m_\pi^{\rm phys}$, 
and not for higher values of $m_\pi$.
On the other hand, it is well known that in 
lattice QCD simulations with large pion masses, 
the meson cloud effects effects are suppressed.
In these conditions, although
one may expect that the valence quark component 
for $m_\pi \approx 300$ MeV provide a close estimate 
for the valence quark component at the physical limit,
for the meson cloud component one can expect 
a stronger dependence on the pion mass due to chiral effects.

To summarize, 
the deviation between the holographic 
parametrizations from the lattice QCD data 
can be interpreted mainly as a consequence of the suppression of 
the meson cloud effects in the lattice QCD simulations.

To test if the deviation of the holographic model
from the lattice data is in fact the result
of the dominance of the valence quark contribution in the lattice data, 
we compare directly the model parametrizations
for the valence quark contributions  
with the results of the lattice QCD simulations.

To help the discussion, we rewrite Eq.~(\ref{eqGA0}) as
\ba
G_A(Q^2)= G_A^{\rm B} (Q^2) + (1-c_3) \left[G_A^{\rm MC} (Q^2) - G_A^{\rm B}(Q^2) \right]. 
\nonumber \\
\label{eqGA1}
\ea
In the present form, the second term can be seen as 
the alternative representation of the meson cloud contribution,
defined by the difference between $G_A$ and $G_A^{\rm B}$,
when all the  normalization factors are taken into account ($Z_N=c_3$).
Notice that the second term in Eq.~(\ref{eqGA1}) vanishes at $Q^2=0$, 
as a consequence of the normalization of $G_A^{\rm B}$ and $G_A^{\rm MC}$ 
(reduced to $g_A^0$ when $Q^2=0$).

Equation (\ref{eqGA1}) provides also a 
simple illustration of the limit where the system is completely 
dominated by the valence quark component.
In that case  $Z_N \equiv c_3 \to 1$, the second term vanishes, 
and $G_A$ is reduced to $G_A^{\rm B}$, as expected.

\begin{figure}[t]
\vspace{.5cm}
\centerline{
\mbox{
\includegraphics[width=3.0in]{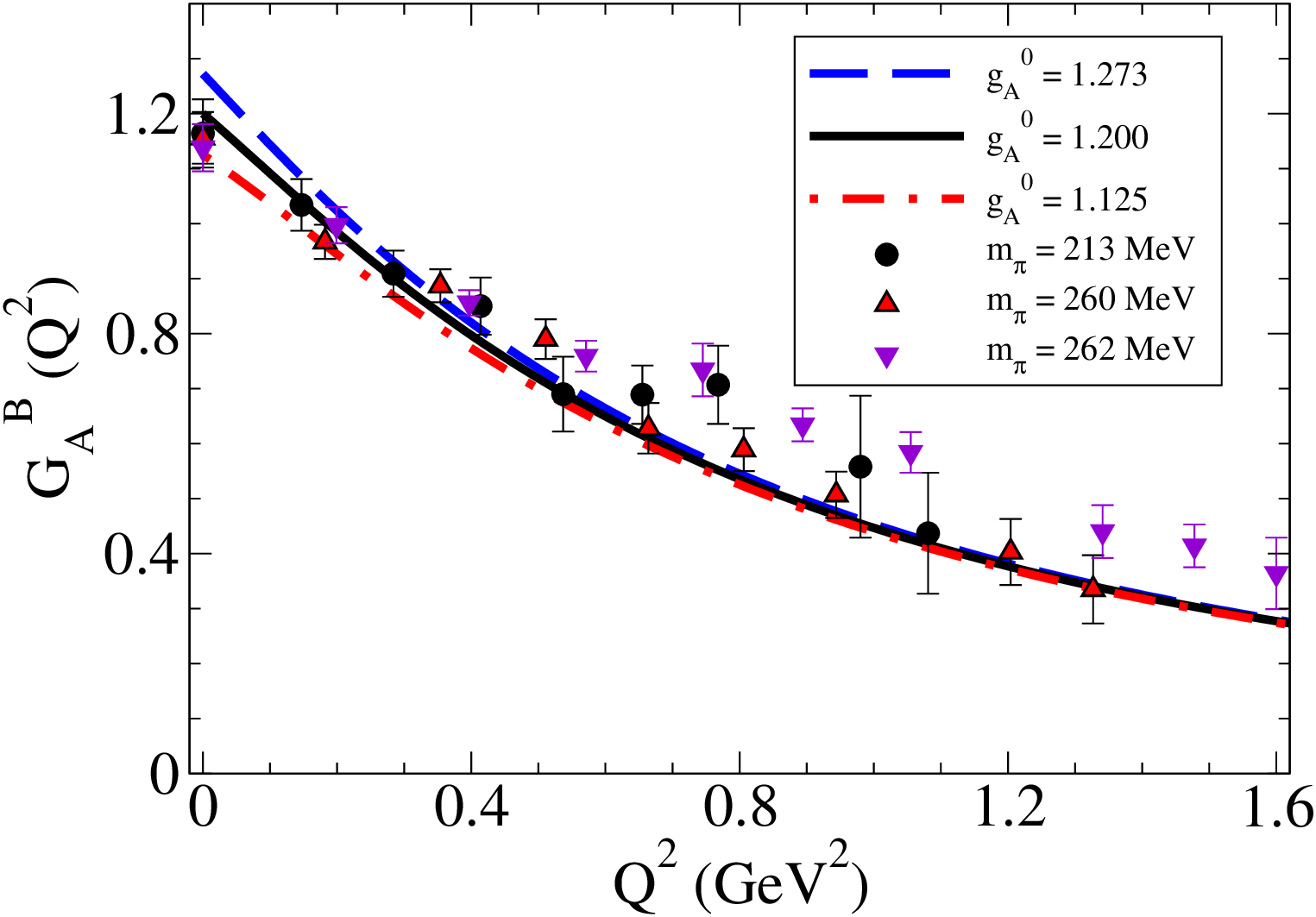} }}
\caption{\footnotesize{
Comparison of $G_A^{\rm B}$ with the
lattice QCD data~\cite{Alexandrou13,Alexandrou11a}.
The models are labeled with the value of 
$g_A^0$ presented in Table~\ref{tableMods}.
}}
%\vspace{-1cm}
\label{figLat}
\end{figure}

The direct comparison between the parametrizations
of the valence quark contribution with the lattice QCD data 
is presented in Fig.~\ref{figLat}.
From the figure, we can conclude that the models with larger $g_A^0$
overestimates the lattice data near $Q^2=0$.
Only the models with the values  $g_A^0=1.125$ and  1.2 are closer 
to the lattice data for $Q^2 < 0.2$ GeV$^2$.
Between those parametrizations, 
$g_A^0=1.2$ is the one that gives the best description 
of the lattice QCD data, as can be observed in Fig.~\ref{figLat}.


Overall, the agreement between the estimate of the 
valence quark contributions and the lattice QCD data 
is better than the previous case, where we compared the full result 
(bare plus meson cloud) with the lattice QCD data.
Notice, in particular the good agreement 
between the estimates at large $Q^2$ 
for the datasets with $m_\pi= 213$ and 260 MeV.
These results suggest that the second term in Eq.~(\ref{eqGA1}),
has a small magnitude in the lattice QCD 
simulations for pion masses around 0.23 GeV.
Notice also that the term under discussion
is negative because  $G_A^{\rm MC}$ has a faster falloff than $G_A^{\rm B}$.
As a consequence, the values of $G_A$ increase, 
when the term is neglected (compare Figs.~\ref{figLat2} and \ref{figLat}).

Looking in particular for the data associated 
with the largest pion mass ($m_\pi = 262$ MeV), 
we  can notice that the function $G_A^B$ 
(estimated for $m_\pi^{\rm phys}$) 
falls off faster with $Q^2$ than the lattice QCD data.
The same effect happens for simulations with  $m_\pi >  300$ MeV
(not shown here). 
This effect has been observed in several lattice QCD studies.
Lattice QCD calculations of form factors 
associated with large pion masses have slower falloffs 
than in the case of the physical form 
factors~\cite{NucleonMC3,Octet2,Lattice,Yamazaki09,Alexandrou13,Sasaki08}.

The differences between the lattice QCD data 
associated with $m_\pi =$ 260 and 262 MeV 
(close values), displayed in Figs.~\ref{figLat2} and \ref{figLat},
suggest that the estimation of 
the valence quark and meson cloud contributions 
for $G_A$ should not be performed based on
only a few lattice QCD datasets.
It is then preferable to use a significant 
number of datasets with different values for the pion masses, 
or in alternative to consider an extrapolation 
of the lattice QCD results based on several datasets, 
as discussed in Sec.~\ref{secLattice}.

The present analysis does not imply that 
the lattice QCD simulations with $m_\pi \approx 0.2$ GeV 
have no meson cloud contributions,
it shows only that those contributions seem to be small or 
of the order of the errorbars. 
Those effects are expected to became more significant 
when we approach the physical limit.

The comparison between the bare contribution of the holographic model 
with the lattice QCD data, and their close agreement, 
justifies the choice of values of $g_A^0$ smaller than 
the experimental value, namely $g_A^0 \simeq 1.2$.
This result confirms also the need to use constraints 
in the function $G_A^{\rm B}$,
in order to obtain a better description of the physics associated 
with the axial form factor $G_A$.

A choice of values of $g_A^0$ below 1.27  may also be justified 
by dynamical effects in the quark structure.
Calculations based on the Dyson-Schwinger framework 
show a reduction of the quark axial charge $g_A^q$ 
due to the gluon dressing of the quarks.
As a consequence the valence quark contribution to $G_A$
is reduced when compared to calculations based on 
undressed quarks~\cite{Eichmann12,Chang13,Yamanaka14}.

\subsection{Vector meson dominance models}

In the literature we can find some models 
for the axial form factor based 
on VMD with axial-vector mass poles~\cite{Hong07,Hashimoto08,Adamuscin08}.
The models from Refs.~\cite{Hong07,Adamuscin08}
are called two-component models,
and include a term associated with the lowest 
axial-vector meson state ($a_1$).
Those models explore also the possible decompositions 
between a bare core component and 
a component associated with the (axial-vector) meson cloud.
The model estimates are compatible with 
the parametrization (\ref{eqGAexp}) below 1 GeV$^2$.
The holographic model from  Ref.~\cite{Hashimoto08}
considers an  expansion in the axial-vector meson poles.
In that case it was shown that the final expression for $G_A$ can also 
be approximated by a dipole, at low $Q^2$.

\subsection{Estimate of the meson cloud contribution from holography}

Finally, we discuss the estimate of the meson cloud contribution 
associated with the our best holographic model ($g_A^0=1.2$).
The meson cloud contribution to the axial form factor 
was already shown in Fig.~\ref{figMod2}.
In that figure we can see, 
looking at the meson cloud contribution (dotted-line) 
that  $G_A^{\rm MC}$ does not fall to zero very fast.
One can also conclude that for large $Q^2$,
the holographic estimate of the bare contribution underestimates 
the result of the spectator model from Ref.~\cite{AxialFF},
defined by the central value of the blue band.

\begin{figure}[t]
\vspace{.5cm}
\centerline{
\mbox{
\includegraphics[width=3.0in]{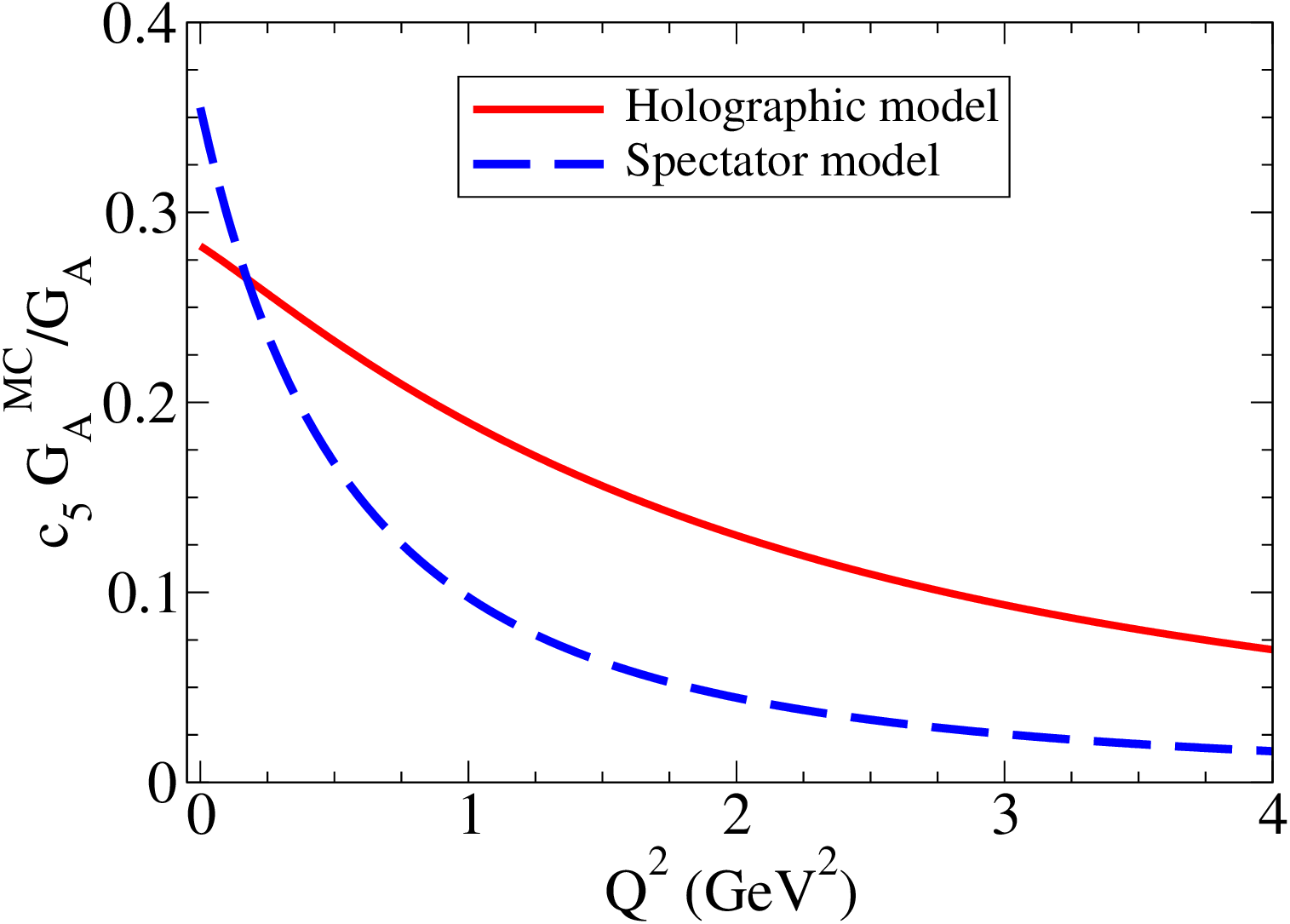} }}
\caption{\footnotesize{
Comparison between the relative contributions 
of the meson cloud in the holographic model and in the spectator model.}}
%\vspace{-1cm}
\label{figMC}
\end{figure}

The previous result suggests that the holographic estimate 
of the meson cloud has a slow falloff
compared to the estimate from the spectator model,
determined by Eq.~(\ref{eqGMCspec})~\cite{AxialFF}.
This effect can be observed in more detail 
in Fig.~\ref{figMC}, where we plot the ratio 
between $c_5 G_A^{\rm MC}$ and $G_A$, estimated by the respective model, 
up to $Q^2=4$ GeV$^2$.
In this representation the difference between falloffs became clear.
Apart the difference between parametrizations 
at $Q^2=0$, which are compatible with 
the uncertainties of the estimates of $G_A^{\rm exp}$ and $G_A^{\rm B}$,
it is clear in the graph the 
difference of falloffs between 
the spectator model (fast) and our best holographic model (slow).

We recall that both estimates of the meson cloud 
fall off with $1/Q^8$ for large $Q^2$
(faster than the valence quark contributions: $1/Q^4$).
The multiplicative factors associated with
those functions in the holographic and spectator models are,  however, very different.
The factor associated with the  holographic model
is larger than the one from the spectator model.

A quantitative measure of the falloff from the   
meson cloud contribution may be the value of $Q^2$ for which 
the contribution of the meson cloud becomes smaller
than 10\% of $G_A(Q^2)$.
From Fig.~\ref{figMC}, we can conclude that this value 
is about 1 GeV$^2$ for the spectator model, 
and about 2.8 GeV$^2$ for the holographic model.

The meson cloud estimate from 
the perturbative chiral quark model 
has a falloff even slower than the holographic model.
For $Q^2 \simeq 1$ GeV$^2$ the meson cloud contribution 
dominates over the bare contribution~\cite{Liu15a,Liu16a}.
According to Ref.~\cite{Liu16a} the flat behavior 
of the meson cloud contribution (slow falloff) 
may indicate that the meson cloud distribution is closer to 
the origin in the coordinate space, than in other models.
The calculation based on the holographic model,
and the faster falloff of the meson cloud contribution, 
suggests a much more peripheral distribution 
of the meson cloud.

The difference between the 
falloffs in holographic models and quark models 
may be a consequence of the way the meson structure is described.
In the holographic models the substructure associated 
with the $q \bar q$ pair is neglected in first approximation, 
meaning that the meson states are regarded as pointlike particles.
In the quark models, the mesons are extended particles 
with structure form factors that can be approximated by 
multipole functions.
Those multipole functions are 
parametrized by cutoffs that characterize 
the spatial extension of the mesons
and are also responsible for the faster falloff 
of the meson cloud contribution, 
compared to models with pointlike mesons.

%\clearpage

\section{Outlook and conclusions}
\label{secConclusions}

In the present work we study the structure 
of the nucleon axial form factor 
using the formalism of the light-front holography.
In a holographic model the substructure 
associated with the valence quark degrees of freedom 
and the substructure associated with the meson cloud excitations  
are both parametrized in terms of two independent microscopic couplings:
$g_A^0$, the quark axial-vector coupling,
and $\eta_A$ the quark induced pseudoscalar coupling.
Contrary to the case of the quark models with meson cloud dressing,
in the holographic models 
there is no explicit connection with the baryon-meson substructure.

We checked if the empirical information 
associated with the nucleon axial form factor $G_A$
could be used to determine the fraction of $G_A$
associated with the valence quark components ($G_A^{\rm B}$)
and the fraction associated with the meson cloud component  ($G_A^{\rm MC}$),
based on a holographic model.
We concluded that this goal can be achieved  
if we use the information from lattice QCD simulations
to constraint the bare component, 
associated with the valence quark degrees of freedom.

We realized also that the results from 
the bare and the meson cloud components of $G_A$
depend crucially on the coupling $g_A^0$.
Large values of $g_A^0$ fail to describe the 
magnitude of the function $G_A^{\rm B}$, extracted from lattice QCD.
Small values of $g_A^0$ fail to describe the empirical data.

A good compromise in the description 
of the experimental data and the estimate of $G_A^{\rm B}$ 
is obtained when we use $g_A^0 \simeq 1.2$,
$\eta_A \simeq 1.1$ and a meson cloud mixture of about 30\% 
in the physical  nucleon  state.
A holographic model with $g_A^0 \simeq 1.2$,
provides also a parametrization more consistent 
with the results from lattice QCD.
Most of lattice QCD simulations give $G_A(0) =1.1$--1.2,
in a wide range of pion masses  ($m_\pi =0.2$--0.5 GeV).

To summarize, the holographic model presented here 
provides a consistent description of the $G_A$ data 
and from the estimate of the bare contribution extracted from lattice QCD.
In addition, the holographic model provides 
a parametrization for the meson cloud contributions to $G_A$.

The holographic estimate of $G_A^{\rm MC}$ has a very slow falloff with $Q^2$.
The meson cloud contribution is  smaller 
than 10\% of $G_A$ only for large values of $Q^2$ ($Q^2 > 2.8$ GeV$^2$).
In other quark models with meson cloud dressing  
this reduction happens typically for values of $Q^2$
larger than 1 GeV$^2$.

In the future, it  will be very interesting to check
if also for the electromagnetic form factors 
estimated by holographic models,
the falloff of the meson cloud contribution 
is very slow as for the nucleon axial form $G_A$, 
or if the falloff is faster, as suggested by some quark models.

\newpage

%\vspace{.3cm}
\begin{acknowledgments}
\vspace{-.4cm}
This work was supported by the
Universidade Federal do
Rio Grande do Norte/Minist\'erio da Educa\c{c}\~ao (UFRN/MEC)
and partially supported by the Funda\c{c}\~ao de Amparo \`a 
Pesquisa do Estado de S\~ao Paulo (FAPESP):
project no.~2017/02684-5, grant no.~2017/17020-BCO-JP. 
\end{acknowledgments}


\begin{thebibliography}{00}

% AdS/QCD references

\bibitem{Maldacena99} 
  J.~M.~Maldacena,
  %``The Large N limit of superconformal field theories and supergravity,''
  Int.\ J.\ Theor.\ Phys.\  {\bf 38}, 1113 (1999)
  [Adv.\ Theor.\ Math.\ Phys.\  {\bf 2}, 231 (1998)]
  %% doi:10.1023/A:1026654312961
  [hep-th/9711200].
  %%CITATION = doi:10.1023/A:1026654312961;%%
  %12593 citations counted in INSPIRE as of 15 Mar 2017

\bibitem{Witten98} 
  E.~Witten,
  %``Anti-de Sitter space and holography,''
  Adv.\ Theor.\ Math.\ Phys.\  {\bf 2}, 253 (1998)
  [hep-th/9802150].
  %%CITATION = HEP-TH/9802150;%%
  %8273 citations counted in INSPIRE as of 15 Mar 2017

\bibitem{Gubser98} 
  S.~S.~Gubser, I.~R.~Klebanov and A.~M.~Polyakov,
  %``Gauge theory correlators from noncritical string theory,''
  Phys.\ Lett.\ B {\bf 428}, 105 (1998)
  %% doi:10.1016/S0370-2693(98)00377-3
  [hep-th/9802109].
  %%CITATION = doi:10.1016/S0370-2693(98)00377-3;%%
  %7128 citations counted in INSPIRE as of 15 Mar 2017


\bibitem{Brodsky15}  % BRODSKY - review 
  S.~J.~Brodsky, G.~F.~de Teramond, H.~G.~Dosch and J.~Erlich,
  %``Light-Front Holographic QCD and Emerging Confinement,''
  Phys.\ Rept.\  {\bf 584}, 1 (2015)
  %% doi:10.1016/j.physrep.2015.05.001
  [arXiv:1407.8131 [hep-ph]].
  %%CITATION = doi:10.1016/j.physrep.2015.05.001;%%
  %96 citations counted in INSPIRE as of 17 Oct 2016


\bibitem{Brodsky08a} 
  S.~J.~Brodsky and G.~F.~de Teramond,
  %``Light-Front Dynamics and AdS/QCD Correspondence: The Pion Form Factor in the Space- and Time-Like Regions,''
  Phys.\ Rev.\ D {\bf 77}, 056007 (2008)
  %% doi:10.1103/PhysRevD.77.056007
  [arXiv:0707.3859 [hep-ph]].
  %%CITATION = doi:10.1103/PhysRevD.77.056007;%%
  %281 citations counted in INSPIRE as of 02 Mar 2016
  %% PION FF


\bibitem{Teramond09a} 
  G.~F.~de Teramond and S.~J.~Brodsky,
  %``Light-Front Holography: A First Approximation to QCD,''
  Phys.\ Rev.\ Lett.\  {\bf 102}, 081601 (2009)
  %% doi:10.1103/PhysRevLett.102.081601
  [arXiv:0809.4899 [hep-ph]].
  %%CITATION = doi:10.1103/PhysRevLett.102.081601;%%
  %220 citations counted in INSPIRE as of 13 Dec 2016



% HOLOGRAPHY



\bibitem{Karch06}
  A.~Karch, E.~Katz, D.~T.~Son and M.~A.~Stephanov,
  %``Linear confinement and AdS/QCD,''
  Phys.\ Rev.\ D {\bf 74}, 015005 (2006) 
  %doi:10.1103/PhysRevD.74.015005
  [hep-ph/0602229].
  %%CITATION = doi:10.1103/PhysRevD.74.015005;%%
  %% spectrum


% rho-meson spectrum
\bibitem{Grigoryan07a} 
  H.~R.~Grigoryan and A.~V.~Radyushkin,
  %``Structure of vector mesons in holographic model with linear confinement,''
  Phys.\ Rev.\ D {\bf 76}, 095007 (2007)
  %% doi:10.1103/PhysRevD.76.095007
  [arXiv:0706.1543 [hep-ph]].
  %%CITATION = doi:10.1103/PhysRevD.76.095007;%%
  %118 citations counted in INSPIRE as of 29 Feb 2016




\bibitem{Branz10a} 
  T.~Branz, T.~Gutsche, V.~E.~Lyubovitskij, I.~Schmidt and A.~Vega,
  %``Light and heavy mesons in a soft-wall holographic approach,''
  Phys.\ Rev.\ D {\bf 82}, 074022 (2010)
  %% doi:10.1103/PhysRevD.82.074022
  [arXiv:1008.0268 [hep-ph]].
  %%CITATION = doi:10.1103/PhysRevD.82.074022;%%
  %83 citations counted in INSPIRE as of 19 Aug 2016


\bibitem{Gutsche12b} 
  T.~Gutsche, V.~E.~Lyubovitskij, I.~Schmidt and A.~Vega,
  %``Dilaton in a soft-wall holographic approach to mesons and baryons,''
  Phys.\ Rev.\ D {\bf 85}, 076003 (2012)
  %% doi:10.1103/PhysRevD.85.076003
  [arXiv:1108.0346 [hep-ph]].
  %%CITATION = doi:10.1103/PhysRevD.85.076003;%%
  %76 citations counted in INSPIRE as of 07 Jul 2016




\bibitem{Teramond15a} 
  G.~F.~de Teramond, H.~G.~Dosch and S.~J.~Brodsky,
  %``Baryon Spectrum from Superconformal Quantum Mechanics and its Light-Front Holographic Embedding,''
  Phys.\ Rev.\ D {\bf 91}, 045040 (2015)
  %% doi:10.1103/PhysRevD.91.045040
  [arXiv:1411.5243 [hep-ph]].
  %%CITATION = doi:10.1103/PhysRevD.91.045040;%%
  %18 citations counted in INSPIRE as of 27 Feb 2016
  %% spectrum 








\bibitem{Chakrabarti13a} 
  D.~Chakrabarti and C.~Mondal,
  %``Nucleon and flavor form factors in a light front quark model in AdS/QCD,''
  %%  Nucleon and flavor form factors in AdS/QCD
  Eur.\ Phys.\ J.\ C {\bf 73}, 2671 (2013)
  %% doi:10.1140/epjc/s10052-013-2671-8
  [arXiv:1307.7995 [hep-ph]].
  %%CITATION = doi:10.1140/epjc/s10052-013-2671-8;%%
  %19 citations counted in INSPIRE as of 27 Feb 2016



\bibitem{Abidin09} 
  Z.~Abidin and C.~E.~Carlson,
  %``Nucleon electromagnetic and gravitational form factors from holography,''
  Phys.\ Rev.\ D {\bf 79}, 115003 (2009)
  %% doi:10.1103/PhysRevD.79.115003
  [arXiv:0903.4818 [hep-ph]].
  %%CITATION = doi:10.1103/PhysRevD.79.115003;%%
  %75 citations counted in INSPIRE as of 29 Feb 2016





\bibitem{Teramond11a} 
  G.~F.~de Teramond and S.~J.~Brodsky,
  %``Excited Baryons in Holographic QCD,''
  AIP Conf.\ Proc.\  {\bf 1432}, 168 (2012)
  %% doi:10.1063/1.3701207
  [arXiv:1108.0965 [hep-ph]].
  %%CITATION = doi:10.1063/1.3701207;%%
  %23 citations counted in INSPIRE as of 14 Mar 2017
  %% ROPER


\bibitem{Liu15c} 
  T.~Liu and B.~Q.~Ma,
  %``Baryon properties from light-front holographic QCD,''
  Phys.\ Rev.\ D {\bf 92}, 096003 (2015)
  %% doi:10.1103/PhysRevD.92.096003
  [arXiv:1510.07783 [hep-ph]].
  %%CITATION = doi:10.1103/PhysRevD.92.096003;%%
  %6 citations counted in INSPIRE as of 19 May 2016


\bibitem{Gutsche12a} 
  T.~Gutsche, V.~E.~Lyubovitskij, I.~Schmidt and A.~Vega,
  %``Nucleon structure including high Fock states in AdS/QCD,''
  Phys.\ Rev.\ D {\bf 86}, 036007 (2012)
  %% doi:10.1103/PhysRevD.86.036007
  [arXiv:1204.6612 [hep-ph]].
  %%CITATION = doi:10.1103/PhysRevD.86.036007;%%
  %34 citations counted in INSPIRE as of 29 Feb 2016


\bibitem{Gutsche13a} 
  T.~Gutsche, V.~E.~Lyubovitskij, I.~Schmidt and A.~Vega,
  %``Nucleon resonances in AdS/QCD,''
  Phys.\ Rev.\ D {\bf 87}, 016017 (2013)
  %% doi:10.1103/PhysRevD.87.016017
  [arXiv:1212.6252 [hep-ph]].
  %%CITATION = doi:10.1103/PhysRevD.87.016017;%%
  %20 citations counted in INSPIRE as of 27 Feb 2016
  %%  ROPER



\bibitem{Sufian17a}
  R.~S.~Sufian, G.~F.~de Teramond, S.~J.~Brodsky, A.~Deur and H.~G.~Dosch,
  %``Analysis of nucleon electromagnetic form factors from light-front holographic QCD : The spacelike region,''
  Phys.\ Rev.\ D {\bf 95}, 014011 (2017)
  %%doi:10.1103/PhysRevD.95.014011
  [arXiv:1609.06688 [hep-ph]].
  %%CITATION = doi:10.1103/PhysRevD.95.014011;%%
  %5 citations counted in INSPIRE as of 10 Feb 2017




\bibitem{RoperHol}
  G.~Ramalho and D.~Melnikov,
  %``Valence quark contributions for the $\gamma^\ast N \to N(1440)$ form factors from light-front holography,''
  Phys.\ Rev.\ D {\bf 97}, 034037 (2018)
  %% doi:10.1103/PhysRevD.97.034037
  [arXiv:1703.03819 [hep-ph]].
  %%CITATION = doi:10.1103/PhysRevD.97.034037;%%
  %4 citations counted in INSPIRE as of 15 Mar 2018 
  %% ROPER



\bibitem{RoperAn}
   G.~Ramalho,
  %``Analytic parametrizations of the $\gamma^\ast N \to N(1440)$ form factors inspired by light-front holography,''
  Phys.\ Rev.\ D {\bf 96}, 054021 (2017)
  %% doi:10.1103/PhysRevD.96.054021
  [arXiv:1706.05707 [hep-ph]].
  %%CITATION = doi:10.1103/PhysRevD.96.054021;%%
  %4 citations counted in INSPIRE as of 09 Feb 2018







\bibitem{Vega11} 
  A.~Vega, I.~Schmidt, T.~Gutsche and V.~E.~Lyubovitskij,
  %``Generalized parton distributions in AdS/QCD,''
  Phys.\ Rev.\ D {\bf 83}, 036001 (2011)
  %%doi:10.1103/PhysRevD.83.036001
  [arXiv:1010.2815 [hep-ph]].
  %%CITATION = doi:10.1103/PhysRevD.83.036001;%%
  %69 citations counted in INSPIRE as of 01 Mar 2017


\bibitem{Chakrabarti13} 
  D.~Chakrabarti and C.~Mondal,
  %``Generalized Parton Distributions for the Proton in AdS/QCD,''
  Phys.\ Rev.\ D {\bf 88}, 073006 (2013)
  %% doi:10.1103/PhysRevD.88.073006
  [arXiv:1307.5128 [hep-ph]].
  %%CITATION = doi:10.1103/PhysRevD.88.073006;%%
  %29 citations counted in INSPIRE as of 04 May 2017










\bibitem{NucleonMC1} 
  %\bibitem{Miller02} 
  G.~A.~Miller,
  %``Light front cloudy bag model: Nucleon electromagnetic form-factors,''
  Phys.\ Rev.\ C {\bf 66}, 032201 (2002)
  %% doi:10.1103/PhysRevC.66.032201
  [nucl-th/0207007]
  %%CITATION = doi:10.1103/PhysRevC.66.032201;%%
  %120 citations counted in INSPIRE as of 22 Apr 2017


\bibitem{NucleonMC2} 
  %\bibitem{Octet0} 
  F.~Gross, G.~Ramalho and K.~Tsushima,
  %``Using baryon octet magnetic moments and masses to fix the pion cloud contribution,''
  Phys.\ Lett.\ B {\bf 690}, 183 (2010)
  %% doi:10.1016/j.physletb.2010.05.016
  [arXiv:0910.2171 [hep-ph]].
  %%CITATION = doi:10.1016/j.physletb.2010.05.016;%%
  %24 citations counted in INSPIRE as of 22 Apr 2017


\bibitem{NucleonMC3}
  G.~Ramalho and K.~Tsushima,
  %``Octet baryon electromagnetic form factors in a relativistic quark model,''
  Phys.\ Rev.\ D {\bf 84}, 054014 (2011)
  %%doi:10.1103/PhysRevD.84.054014
  [arXiv:1107.1791 [hep-ph]].
  %%CITATION = doi:10.1103/PhysRevD.84.054014;%%
  %27 citations counted in INSPIRE as of 20 Jun 2017







\bibitem{AxialFF} 
  G.~Ramalho and K.~Tsushima,
  %``Axial form factors of the octet baryons in a covariant quark model,''
  Phys.\ Rev.\ D {\bf 94}, 014001 (2016)
  %% doi:10.1103/PhysRevD.94.014001
  [arXiv:1512.01167 [hep-ph]].







% Axial structure -- Biblo 

% Review  -- ChPT
\bibitem{Bernard02}
  V.~Bernard, L.~Elouadrhiri and U.~G.~Meissner,
  %``Axial structure of the nucleon: Topical Review,''
  J.\ Phys.\ G {\bf 28}, R1 (2002)
  [hep-ph/0107088].
  %%CITATION = HEP-PH/0107088;%%
  %270 citations counted in INSPIRE as of 13 Jan 2015


% Review - proceedings
\bibitem{Schindler07a} 
  M.~R.~Schindler and S.~Scherer,
  %``Nucleon Form Factors of the Isovector Axial-Vector Current:
  % Situation of Experiments and Theory,''
  Eur.\ Phys.\ J.\ A {\bf 32}, 429 (2007)
  [hep-ph/0608325].
  %%CITATION = HEP-PH/0608325;%%
  %12 citations counted in INSPIRE as of 03 juin 2015



% Review -- GP
\bibitem{Gorringe05} 
  T.~Gorringe and H.~W.~Fearing,
  %``Induced pseudoscalar coupling of the proton weak interaction,''
  Rev.\ Mod.\ Phys.\  {\bf 76}, 31 (2003)
  [nucl-th/0206039].
  %%CITATION = NUCL-TH/0206039;%%
  %79 citations counted in INSPIRE as of 03 juin 2015


\bibitem{PDG2014} 
  K.~A.~Olive {\it et al.}  [Particle Data Group Collaboration],
  %``Review of Particle Physics,''
  Chin.\ Phys.\ C {\bf 38}, 090001 (2014).
  %%CITATION = CHPHD,C38,090001;%%
  %1178 citations counted in INSPIRE as of 03 Jun 2015


% ----------------------------------------------------------------------
% QM and others


\bibitem{NSTAR} 
  I.~G.~Aznauryan {\it et al.},
  %``Studies of Nucleon Resonance Structure in Exclusive Meson Electroproduction,''
  Int.\ J.\ Mod.\ Phys.\ E {\bf 22}, 1330015 (2013)
  %% doi:10.1142/S0218301313300154
  [arXiv:1212.4891 [nucl-th]].
  %%CITATION = doi:10.1142/S0218301313300154;%%
  %103 citations counted in INSPIRE as of 20 Apr 2017


% SU(6)

\bibitem{Gaillard84}
  J.~M.~Gaillard and G.~Sauvage,
  %``Hyperon Beta Decays,''
  Ann.\ Rev.\ Nucl.\ Part.\ Sci.\  {\bf 34}, 351 (1984).
  %%CITATION = ARNUA,34,351;%%
  %95 citations counted in INSPIRE as of 13 Jan 2015


\bibitem{JDiaz04} 
  B.~Julia-Diaz, D.~O.~Riska and F.~Coester,
  %``Axial transition form-factors and pion decay of baryon resonances,''
  Phys.\ Rev.\ C {\bf 70}, 045204 (2004)
  %%doi:10.1103/PhysRevC.70.045204
  [nucl-th/0406015].
  %%CITATION = doi:10.1103/PhysRevC.70.045204;%%
  %11 citations counted in INSPIRE as of 26 Apr 2017



\bibitem{Adamuscin08}
  C.~Adamuscin, E.~Tomasi-Gustafsson, E.~Santopinto and R.~Bijker,
  %``Two-component model for the axial form factor of the nucleon,''
  Phys.\ Rev.\ C {\bf 78}, 035201 (2008)
  [arXiv:0706.3509 [nucl-th]].
  %%CITATION = ARXIV:0706.3509;%%
  %3 citations counted in INSPIRE as of 03 Sep 2013
  % Quark model + meson cloud


% CBM - Meson cloud

\bibitem{Thomas84}
  %\bibitem{Thomas84}
  A.~W.~Thomas,
  %``Chiral Symmetry And The Bag Model: A New Starting Point For Nuclear
  %Physics,''
  Adv.\ Nucl.\ Phys.\  {\bf 13}, 1 (1984);
  % discussion of GP
  %\bibitem{Thomas83}
  S.~Theberge and A.~W.~Thomas,
  %``Magnetic Moments Of The Nucleon Octet Calculated In The Cloudy Bag Model,''
  Nucl.\ Phys.\  A {\bf 393}, 252 (1983).
  %%CITATION = NUPHA,A393,252;%%

\bibitem{Tsushima88}
  K.~Tsushima, T.~Yamaguchi, Y.~Kohyama and K.~Kubodera,
  %``Weak Interaction Form-factors And Magnetic Moments Of Octet Baryons: Chiral Bag Model With Gluonic Effects,''
  Nucl.\ Phys.\ A {\bf 489}, 557 (1988).
  %%CITATION = NUPHA,A489,557;%%
  %22 citations counted in INSPIRE as of 05 Sep 2013




\bibitem{Shanahan13}
  P.~E.~Shanahan, A.~W.~Thomas, K.~Tsushima, R.~D.~Young and F.~Myhrer,
  %``Octet Spin Fractions and the Proton Spin Problem,''
  Phys.\ Rev.\ Lett.\  {\bf 110}, 202001 (2013)
  [arXiv:1302.6300 [nucl-th]].
  %%CITATION = ARXIV:1302.6300;%%
  %5 citations counted in INSPIRE as of 20 Jan 2015






%  QM - GA
\bibitem{Schlumpf93a} 
  F.~Schlumpf,
  %``Relativistic constituent quark model of electroweak properties of baryons,''
  Phys.\ Rev.\ D {\bf 47}, 4114 (1993)
  [Phys.\ Rev.\ D {\bf 49}, 6246 (1994)]
  [hep-ph/9212250].
  %%CITATION = HEP-PH/9212250;%%
  %89 citations counted in INSPIRE as of 20 Aug 2015


\bibitem{Boffi02}
  S.~Boffi, L.~Y.~Glozman, W.~Klink, W.~Plessas, M.~Radici and R.~F.~Wagenbrunn,
  %``Covariant electroweak nucleon form-factors in a chiral constituent quark model,''
  Eur.\ Phys.\ J.\ A {\bf 14}, 17 (2002)
  [hep-ph/0108271].
  %%CITATION = HEP-PH/0108271;%%
  %119 citations counted in INSPIRE as of 22 Nov 2013
  %% INCLUDE GP


% RQM
\bibitem{Merten02} 
  D.~Merten, U.~Loring, K.~Kretzschmar, B.~Metsch and H.~R.~Petry,
  %``Electroweak form-factors of nonstrange baryons,''
  Eur.\ Phys.\ J.\ A {\bf 14}, 477 (2002)
  [hep-ph/0204024].
  %%CITATION = HEP-PH/0204024;%%
  %68 citations counted in INSPIRE as of 20 août 2015


% NRQM - ChiQM
\bibitem{BCano03} 
  D.~Barquilla-Cano, A.~J.~Buchmann and E.~Hernandez,
  %``Partial conservation of axial current and axial exchange currents in the nucleon,''
  Nucl.\ Phys.\ A {\bf 714}, 611 (2003)
  [nucl-th/0204067].
  %%CITATION = NUCL-TH/0204067;%%
  %8 citations counted in INSPIRE as of 20 août 2015




% chiral soliton model
\bibitem{Silva05} 
  A.~Silva, H.~C.~Kim, D.~Urbano and K.~Goeke,
  %``Axial-vector form-factors of the nucleon within the chiral quark-soliton model and their strange components,''
  Phys.\ Rev.\ D {\bf 72}, 094011 (2005)
  [hep-ph/0509281].
  %%CITATION = HEP-PH/0509281;%%
  %23 citations counted in INSPIRE as of 20 août 2015


\bibitem{Pasquini07}
  B.~Pasquini and S.~Boffi,
  %``Electroweak structure of the nucleon, meson cloud and light-cone wavefunctions,''
  Phys.\ Rev.\ D {\bf 76}, 074011 (2007)
  [arXiv:0707.2897 [hep-ph]].
  %%CITATION = ARXIV:0707.2897;%%
  %38 citations counted in INSPIRE as of 13 Jan 2015





\bibitem{Dahiya14a} 
  H.~Dahiya and M.~Randhawa,
  %``Axial-vector form factors for the low lying octet baryons in the chiral quark constituent model,''
  Phys.\ Rev.\ D {\bf 90}, 074001 (2014)
  %%doi:10.1103/PhysRevD.90.074001
  [arXiv:1409.4943 [hep-ph]].
  %%CITATION = doi:10.1103/PhysRevD.90.074001;%%
  %9 citations counted in INSPIRE as of 04 May 2017




\bibitem{Anikin16a} 
  I.~V.~Anikin, V.~M.~Braun and N.~Offen,
  %``Axial form factor of the nucleon at large momentum transfers,''
  Phys.\ Rev.\ D {\bf 94}, 034011 (2016)
  %% doi:10.1103/PhysRevD.94.034011
  [arXiv:1607.01504 [hep-ph]].
  %%CITATION = doi:10.1103/PhysRevD.94.034011;%%
  %2 citations counted in INSPIRE as of 04 May 2017





\bibitem{Liu15a} 
  X.~Y.~Liu, K.~Khosonthongkee, A.~Limphirat, P.~Suebka and Y.~Yan,
  %``Meson cloud contributions to baryon axial form factors,''
  Phys.\ Rev.\ D {\bf 91}, 034022 (2015)
  %%doi:10.1103/PhysRevD.91.034022
  [arXiv:1406.7633 [hep-ph]].
  %%CITATION = doi:10.1103/PhysRevD.91.034022;%%
  %3 citations counted in INSPIRE as of 25 Apr 2017

\bibitem{Liu16a} 
  X.~Y.~Liu, K.~Khosonthongkee, A.~Limphirat and Y.~Yan,
  %``Comparisons of electric charge and axial charge meson cloud distributions in the PCQM,''
  arXiv:1601.01428 [hep-ph].
  %%CITATION = ARXIV:1601.01428;%%





% Dyson-Schwinger

\bibitem{Eichmann12}
  G.~Eichmann and C.~S.~Fischer,
  %``Nucleon axial and pseudoscalar form factors from the covariant Faddeev equation,''
  Eur.\ Phys.\ J.\ A {\bf 48}, 9 (2012)
  [arXiv:1111.2614 [hep-ph]].
  %%CITATION = ARXIV:1111.2614;%%
  %23 citations counted in INSPIRE as of 13 Jan 2015
  % Include GP


\bibitem{Chang13}
  L.~Chang, C.~D.~Roberts and S.~M.~Schmidt,
  %``Dressed-quarks and the nucleon's axial charge,''
  Phys.\ Rev.\ C {\bf 87}, 015203 (2013)
  [arXiv:1207.5300 [nucl-th]].
  %%CITATION = ARXIV:1207.5300;%%
  %12 citations counted in INSPIRE as of 13 Jan 2015


\bibitem{Yamanaka14} 
  N.~Yamanaka, S.~Imai, T.~M.~Doi and H.~Suganuma,
  %``Quark scalar, axial, and pseudoscalar charges in the Schwinger-Dyson formalism,''
  Phys.\ Rev.\ D {\bf 89}, 074017 (2014)
  [arXiv:1401.2852 [hep-ph]].
  %%CITATION = ARXIV:1401.2852;%%
  %6 citations counted in INSPIRE as of 28 Jul 2015





\bibitem{Mamedov16a} 
  S.~Mamedov, B.~B.~Sirvanli, I.~Atayev and N.~Huseynova,
  %``Nucleon's axial-vector form factor in the hard-wall AdS/QCD model,''
  Int.~J.~Theor.~Phys.~(2017)
  %% doi:10.1007/s10773-017-3330-x
  arXiv:1609.00167 [hep-th].
  %%CITATION = doi:10.1007/s10773-017-3330-x;%%
  %1 citations counted in INSPIRE as of 03 May 2017










% Lattice biblo  -----------------------------------------------------

% Q2=0

\bibitem{Sasaki03} 
  S.~Sasaki {\it et al.} [RIKEN-BNL-Columbia-KEK Collaboration],
  %``Nucleon axial charge from quenched lattice QCD with domain wall fermions,''
  Phys.\ Rev.\ D {\bf 68}, 054509 (2003)
  %%doi:10.1103/PhysRevD.68.054509
  [hep-lat/0306007].
  %%CITATION = doi:10.1103/PhysRevD.68.054509;%%
  %77 citations counted in INSPIRE as of 27 Apr 2017

\bibitem{Edwards06} 
  R.~G.~Edwards {\it et al.} [LHPC Collaboration],
  %``The Nucleon axial charge in full lattice QCD,''
  Phys.\ Rev.\ Lett.\  {\bf 96}, 052001 (2006)
  %% doi:10.1103/PhysRevLett.96.052001
  [hep-lat/0510062].
  %%CITATION = doi:10.1103/PhysRevLett.96.052001;%%
  %153 citations counted in INSPIRE as of 27 Apr 2017


\bibitem{Yamazaki08} 
  T.~Yamazaki {\it et al.} [RBC+UKQCD Collaboration],
  %``Nucleon axial charge in 2+1 flavor dynamical lattice QCD with domain wall fermions,''
  Phys.\ Rev.\ Lett.\  {\bf 100}, 171602 (2008)
  %% doi:10.1103/PhysRevLett.100.171602
  [arXiv:0801.4016 [hep-lat]].
  %%CITATION = doi:10.1103/PhysRevLett.100.171602;%%
  %103 citations counted in INSPIRE as of 27 Apr 2017




\bibitem{Bhattacharya14}  
  T.~Bhattacharya, S.~D.~Cohen, R.~Gupta, A.~Joseph, H.~W.~Lin and B.~Yoon,
  %``Nucleon Charges and Electromagnetic Form Factors from 2+1+1-Flavor Lattice QCD,''
  Phys.\ Rev.\ D {\bf 89}, 094502 (2014)
  %% doi:10.1103/PhysRevD.89.094502
  [arXiv:1306.5435 [hep-lat]].
  %%CITATION = doi:10.1103/PhysRevD.89.094502;%%
  %89 citations counted in INSPIRE as of 27 Apr 2017

\bibitem{Liang16} 
  J.~Liang, Y.~B.~Yang, K.~F.~Liu, A.~Alexandru, T.~Draper and R.~S.~Sufian,
  %``Lattice Calculation of Nucleon Isovector Axial Charge with Improved Currents,''
  arXiv:1612.04388 [hep-lat].
  %%CITATION = ARXIV:1612.04388;%%
  %6 citations counted in INSPIRE as of 12 Jul 2017




% Lattice bibliography Q2 > 0 

\bibitem{Sasaki08} 
  S.~Sasaki and T.~Yamazaki,
  %``Nucleon form factors from quenched lattice QCD with domain wall fermions,''
  Phys.\ Rev.\ D {\bf 78}, 014510 (2008)
  [arXiv:0709.3150 [hep-lat]].
  %%CITATION = ARXIV:0709.3150;%%
  %20 citations counted in INSPIRE as of 08 sept. 2015
  %% Include GP


\bibitem{Yamazaki09} 
  T.~Yamazaki, Y.~Aoki, T.~Blum, H.~W.~Lin, S.~Ohta, S.~Sasaki, R.~Tweedie and J.~Zanotti,
  %``Nucleon form factors with 2+1 flavor dynamical domain-wall fermions,''
  Phys.\ Rev.\ D {\bf 79}, 114505 (2009)
  %% doi:10.1103/PhysRevD.79.114505
  [arXiv:0904.2039 [hep-lat]].
  %%CITATION = doi:10.1103/PhysRevD.79.114505;%%
  %131 citations counted in INSPIRE as of 27 Apr 2017




\bibitem{Bratt10} 
  J.~D.~Bratt {\it et al.} [LHPC Collaboration],
  %``Nucleon structure from mixed action calculations using 2+1 flavors of asqtad sea and domain wall valence fermions,''
  Phys.\ Rev.\ D {\bf 82}, 094502 (2010)
  %% doi:10.1103/PhysRevD.82.094502
  [arXiv:1001.3620 [hep-lat]].
  %%CITATION = doi:10.1103/PhysRevD.82.094502;%%
  %183 citations counted in INSPIRE as of 27 Apr 2017


\bibitem{Alexandrou13}
  C.~Alexandrou, M.~Constantinou, S.~Dinter, V.~Drach, K.~Jansen, C.~Kallidonis and G.~Koutsou,
  %``Nucleon form factors and moments of generalized parton distributions using $N_f=2+1+1$ twisted mass fermions,''
  Phys.\ Rev.\ D {\bf 88}, 014509 (2013)
  [arXiv:1303.5979 [hep-lat]].
  %%CITATION = ARXIV:1303.5979;%%
  %2 citations counted in INSPIRE as of 03 Sep 2013


\bibitem{Alexandrou11a}
  C.~Alexandrou {\it et al.}  [ETM Collaboration],
  %``Axial Nucleon form factors from lattice QCD,''
  Phys.\ Rev.\ D {\bf 83}, 045010 (2011)
  [arXiv:1012.0857 [hep-lat]].
  %%CITATION = ARXIV:1012.0857;%%
  %53 citations counted in INSPIRE as of 12 Jan 2015


\bibitem{Green17} 
  J.~Green {\it et al.},
  %``Up, down, and strange nucleon axial form factors from lattice QCD,''
  Phys.\ Rev.\ D {\bf 95}, 114502 (2017)
  %% doi:10.1103/PhysRevD.95.114502
  [arXiv:1703.06703 [hep-lat]].
  %%CITATION = doi:10.1103/PhysRevD.95.114502;%%
  %3 citations counted in INSPIRE as of 12 Jul 2017



% Near the physical point

\bibitem{Horsley14}   
  R.~Horsley, Y.~Nakamura, A.~Nobile, P.~E.~L.~Rakow, G.~Schierholz and J.~M.~Zanotti,
  %``Nucleon axial charge and pion decay constant from two-flavor lattice QCD,''
  Phys.\ Lett.\ B {\bf 732}, 41 (2014)
  %% doi:10.1016/j.physletb.2014.03.002
  [arXiv:1302.2233 [hep-lat]].
  %%CITATION = doi:10.1016/j.physletb.2014.03.002;%%
  %44 citations counted in INSPIRE as of 27 Apr 2017


\bibitem{ARehim15}
  A.~Abdel-Rehim {\it et al.},
  %``Nucleon and pion structure with lattice QCD simulations at physical value of the pion mass,''
  Phys.\ Rev.\ D {\bf 92},  114513, (2015)
   Erratum: [Phys.\ Rev.\ D {\bf 93},  039904 (2016)]
  %% doi:10.1103/PhysRevD.92.114513, 10.1103/PhysRevD.93.039904
  [arXiv:1507.04936 [hep-lat]].
  %%CITATION = doi:10.1103/PhysRevD.92.114513, 10.1103/PhysRevD.93.039904;%%
  %58 citations counted in INSPIRE as of 26 Apr 2017


\bibitem{Bali15} 
  G.~S.~Bali {\it et al.},
  %``Nucleon isovector couplings from $N_f=2$ lattice QCD,''
  Phys.\ Rev.\ D {\bf 91}, 054501 (2015)
  %% doi:10.1103/PhysRevD.91.054501
  [arXiv:1412.7336 [hep-lat]].
  %%CITATION = doi:10.1103/PhysRevD.91.054501;%%
  %47 citations counted in INSPIRE as of 27 Apr 2017

\bibitem{Bhattacharya16}
  T.~Bhattacharya, V.~Cirigliano, S.~Cohen, R.~Gupta, H.~W.~Lin and B.~Yoon,
  %``Axial, Scalar and Tensor Charges of the Nucleon from 2+1+1-flavor Lattice QCD,''
  Phys.\ Rev.\ D {\bf 94}, 054508 (2016)
  %% doi:10.1103/PhysRevD.94.054508
  [arXiv:1606.07049 [hep-lat]].
  %%CITATION = doi:10.1103/PhysRevD.94.054508;%%
  %28 citations counted in INSPIRE as of 27 Apr 2017

\bibitem{Capitani17} 
  S.~Capitani {\it et al.},
  %``Iso-vector axial form factors of the nucleon in two-flavour lattice QCD,''
  arXiv:1705.06186 [hep-lat].
  %%CITATION = ARXIV:1705.06186;%%



\bibitem{Berkowitz17} 
  E.~Berkowitz {\it et al.},
  %``An accurate calculation of the nucleon axial charge with lattice QCD,''
  arXiv:1704.01114 [hep-lat].
  %%CITATION = ARXIV:1704.01114;%%
  %9 citations counted in INSPIRE as of 13 Sep 2017


\bibitem{Yao17} 
  D.~L.~Yao, L.~Alvarez-Ruso and M.~J.~Vicente-Vacas,
  %``Extraction of nucleon axial charge and radius from lattice QCD results using baryon chiral perturbation theory,''
  arXiv:1708.08776 [hep-ph].
  %%CITATION = ARXIV:1708.08776;%%






% ============================================================================


\bibitem{Park12} 
  K.~Park {\it et al.} [CLAS Collaboration],
  %``Measurement of the generalized form factors near threshold via $\gamma^* p \to n\pi^+$ at high $Q^2$,''
  Phys.\ Rev.\ C {\bf 85}, 035208 (2012)
  %% doi:10.1103/PhysRevC.85.035208
  [arXiv:1201.0903 [nucl-ex]].
  %%CITATION = doi:10.1103/PhysRevC.85.035208;%%
  %7 citations counted in INSPIRE as of 03 May 2017



% =============================================================================






\bibitem{Nucleon} 
  F.~Gross, G.~Ramalho and M.~T.~Pe\~na,
  %``A Pure S-wave covariant model for the nucleon,''
  Phys.\ Rev.\ C {\bf 77}, 015202 (2008)
  %% doi:10.1103/PhysRevC.77.015202
  [nucl-th/0606029].
  %%CITATION = doi:10.1103/PhysRevC.77.015202;%%
  %75 citations counted in INSPIRE as of 20 Apr 2017



\bibitem{Omega1} 
  G.~Ramalho, K.~Tsushima and F.~Gross,
  %``A Relativistic quark model for the Omega- electromagnetic form factors,''
  Phys.\ Rev.\ D {\bf 80}, 033004 (2009)
  %% doi:10.1103/PhysRevD.80.033004
  [arXiv:0907.1060 [hep-ph]].
  %%CITATION = doi:10.1103/PhysRevD.80.033004;%%
  %39 citations counted in INSPIRE as of 20 Apr 2017


\bibitem{Omega2} 
  G.~Ramalho and M.~T.~Pe\~na,
  %``Extracting the Omega- electric quadrupole moment from lattice QCD data,''
  Phys.\ Rev.\ D {\bf 83}, 054011 (2011)
  %% doi:10.1103/PhysRevD.83.054011
  [arXiv:1012.2168 [hep-ph]].
  %%CITATION = doi:10.1103/PhysRevD.83.054011;%%
  %21 citations counted in INSPIRE as of 20 Apr 2017







\bibitem{Lattice} 
  G.~Ramalho and M.~T.~Pe\~na,
  %``Nucleon and gamma N ---> Delta lattice form factors in a constituent quark model,''
  J.\ Phys.\ G {\bf 36}, 115011 (2009)
  %% doi:10.1088/0954-3899/36/11/115011
  [arXiv:0812.0187 [hep-ph]].
  %%CITATION = doi:10.1088/0954-3899/36/11/115011;%%
  %36 citations counted in INSPIRE as of 20 Apr 2017



\bibitem{LatticeD} 
  G.~Ramalho and M.~T.~Pe\~na,
  %``Valence quark contribution for the gamma N ---> Delta quadrupole transition extracted from lattice QCD,''
  Phys.\ Rev.\ D {\bf 80}, 013008 (2009)
  %% doi:10.1103/PhysRevD.80.013008
  [arXiv:0901.4310 [hep-ph]].
  %%CITATION = doi:10.1103/PhysRevD.80.013008;%%
  %41 citations counted in INSPIRE as of 20 Apr 2017



\bibitem{Octet2} 
  G.~Ramalho, K.~Tsushima and A.~W.~Thomas,
  %``Octet Baryon Electromagnetic form Factors in Nuclear Medium,''
  J.\ Phys.\ G {\bf 40}, 015102 (2013)
  %% doi:10.1088/0954-3899/40/1/015102
  [arXiv:1206.2207 [hep-ph]].
  %%CITATION = doi:10.1088/0954-3899/40/1/015102;%%
  %25 citations counted in INSPIRE as of 20 Apr 2017




%  Top-down approach


\bibitem{Sakai05}
  T.~Sakai and S.~Sugimoto,
  %``Low energy hadron physics in holographic QCD,''
  Prog.\ Theor.\ Phys.\  {\bf 113}, 843 (2005) 
  %doi:10.1143/PTP.113.843
  [hep-th/0412141].
  %%CITATION = doi:10.1143/PTP.113.843;%%


\bibitem{Nawa07} 
  K.~Nawa, H.~Suganuma and T.~Kojo,
  %``Baryons in holographic QCD,''
  Phys.\ Rev.\ D {\bf 75}, 086003 (2007)
  %%doi:10.1103/PhysRevD.75.086003
  [hep-th/0612187].
  %%CITATION = doi:10.1103/PhysRevD.75.086003;%%
  %98 citations counted in INSPIRE as of 08 Feb 2018



\bibitem{Hong07} 
  D.~K.~Hong, M.~Rho, H.~U.~Yee and P.~Yi,
  %``Dynamics of baryons from string theory and vector dominance,''
  JHEP {\bf 0709}, 063 (2007)
  %%doi:10.1088/1126-6708/2007/09/063
  [arXiv:0705.2632 [hep-th]].
  %%CITATION = doi:10.1088/1126-6708/2007/09/063;%%
  %124 citations counted in INSPIRE as of 14 Sep 2017




\bibitem{Hashimoto08} 
  K.~Hashimoto, T.~Sakai and S.~Sugimoto,
  %``Holographic Baryons: Static Properties and Form Factors from Gauge/String Duality,''
  Prog.\ Theor.\ Phys.\  {\bf 120}, 1093 (2008)
  %%doi:10.1143/PTP.120.1093
  [arXiv:0806.3122 [hep-th]].
  %%CITATION = doi:10.1143/PTP.120.1093;%%
  %135 citations counted in INSPIRE as of 14 Sep 2017




\bibitem{Hong08} 
  D.~K.~Hong, M.~Rho, H.~U.~Yee and P.~Yi,
  %``Nucleon form-factors and hidden symmetry in holographic QCD,''
  Phys.\ Rev.\ D {\bf 77}, 014030 (2008)
  %% doi:10.1103/PhysRevD.77.014030
  [arXiv:0710.4615 [hep-ph]].
  %%CITATION = doi:10.1103/PhysRevD.77.014030;%%
  %83 citations counted in INSPIRE as of 14 Sep 2017




\bibitem{Sakurai60} 
  J.~J.~Sakurai,
  %``Theory of strong interactions,''
  Annals Phys.\  {\bf 11}, 1 (1960).
  %% doi:10.1016/0003-4916(60)90126-3
  %%CITATION = doi:10.1016/0003-4916(60)90126-3;%%
  %862 citations counted in INSPIRE as of 12 Jun 2017


\bibitem{Bauer79} 
  T.~H.~Bauer, R.~D.~Spital, D.~R.~Yennie and F.~M.~Pipkin,
  %``The Hadronic Properties of the Photon in High-Energy Interactions,''
  Rev.\ Mod.\ Phys.\  {\bf 50}, 261 (1978)
  Erratum: [Rev.\ Mod.\ Phys.\  {\bf 51}, 407 (1979)].
  %% doi:10.1103/RevModPhys.50.261
  %%CITATION = doi:10.1103/RevModPhys.50.261;%%
  %866 citations counted in INSPIRE as of 15 Jun 2017


\bibitem{Gari84} 
  M.~Gari and W.~Krumpelmann,
  %``Generalized Vector Meson Dominance And The Electromagnetic Structure Of The Nucleon,''
  Phys.\ Lett.\  {\bf 141B}, 295 (1984).
  %% doi:10.1016/0370-2693(84)90248-X
  %%CITATION = doi:10.1016/0370-2693(84)90248-X;%%
  %31 citations counted in INSPIRE as of 12 Jun 2017

\bibitem{Lomon01} 
  E.~L.~Lomon,
  %``Extended Gari-Krumpelmann model fits to nucleon electromagnetic form-factors,''
  Phys.\ Rev.\ C {\bf 64}, 035204 (2001)
  %%doi:10.1103/PhysRevC.64.035204
  [nucl-th/0104039].
  %%CITATION = doi:10.1103/PhysRevC.64.035204;%%
  %94 citations counted in INSPIRE as of 12 Jun 2017

\end{thebibliography}
\end{document}